\documentclass[preprint]{emulateapj-rtx4}

\shorttitle{Subaru High-$z$ Exploration of Low-Luminosity Quasars I}
\shortauthors{Matsuoka et al.}

\begin{document}


\title{Subaru high-{\scriptsize $z$} exploration of low-luminosity quasars (SHELLQs).\\ I.
    Discovery of 15 quasars and bright galaxies at $5.7 < z < 6.9$$^\dagger$$^\ddagger$}


\author{Yoshiki Matsuoka\altaffilmark{1, 2, 3, $\star$}, Masafusa Onoue\altaffilmark{2}, Nobunari Kashikawa\altaffilmark{1, 2}, Kazushi Iwasawa\altaffilmark{4}, 
Michael A. Strauss\altaffilmark{3}, Tohru Nagao\altaffilmark{5}, Masatoshi Imanishi\altaffilmark{1, 2, 6}, Mana Niida\altaffilmark{7}, Yoshiki Toba\altaffilmark{5}, 
Masayuki Akiyama\altaffilmark{8}, Naoko Asami\altaffilmark{9}, James Bosch\altaffilmark{3}, S\'{e}bastien Foucaud\altaffilmark{10}, Hisanori Furusawa\altaffilmark{1}, 
Tomotsugu Goto\altaffilmark{11}, James E. Gunn\altaffilmark{3},
Yuichi Harikane\altaffilmark{12, 13}, Hiroyuki Ikeda\altaffilmark{1}, Toshihiro Kawaguchi\altaffilmark{14}, Satoshi Kikuta\altaffilmark{2}, Yutaka Komiyama\altaffilmark{1, 2}, 
Robert H. Lupton\altaffilmark{3}, Takeo Minezaki\altaffilmark{15}, Satoshi Miyazaki\altaffilmark{1, 2}, Tomoki Morokuma\altaffilmark{15}, Hitoshi Murayama\altaffilmark{16},
Atsushi J. Nishizawa\altaffilmark{17}, Yoshiaki Ono\altaffilmark{12}, 
Masami Ouchi\altaffilmark{12, 16}, Paul A. Price\altaffilmark{3}, Hiroaki Sameshima\altaffilmark{18}, John D. Silverman\altaffilmark{16}, 
Naoshi Sugiyama\altaffilmark{16, 19}, Philip J. Tait\altaffilmark{6}, 
Masahiro Takada\altaffilmark{16}, Tadafumi Takata\altaffilmark{1, 2}, Masayuki Tanaka\altaffilmark{1, 2}, Ji-Jia Tang\altaffilmark{20}, Yousuke Utsumi\altaffilmark{21}}


\altaffiltext{1}{National Astronomical Observatory of Japan, Mitaka, Tokyo 181-8588, Japan.}
\altaffiltext{2}{Department of Astronomy, School of Science, Graduate University for Advanced Studies, Mitaka, Tokyo 181-8588, Japan.}
\altaffiltext{3}{Princeton University Observatory, Peyton Hall, Princeton, NJ 08544, USA.}
\altaffiltext{4}{ICREA and Institut de Ci{\`e}ncies del Cosmos, Universitat de Barcelona, IEEC-UB, Mart{\'i} i Franqu{\`e}s, 1, 08028 Barcelona, Spain.}
\altaffiltext{5}{Research Center for Space and Cosmic Evolution, Ehime University, Matsuyama, Ehime 790-8577, Japan.}
\altaffiltext{6}{Subaru Telescope, Hilo, HI 96720, USA.}
\altaffiltext{7}{Graduate School of Science and Engineering, Ehime University, Matsuyama, Ehime 790-8577, Japan.}
\altaffiltext{8}{Astronomical Institute, Tohoku University, Aoba, Sendai, 980-8578, Japan.}
\altaffiltext{9}{Japan Professional School of Education, Chiyoda, Tokyo 101-0041, Japan.}
\altaffiltext{10}{Department of Physics and Astronomy, Shanghai JiaoTong University, Shanghai 200240, China.}
\altaffiltext{11}{Institute of Astronomy and Department of Physics, National Tsing Hua University, Hsinchu 30013, Taiwan.}
\altaffiltext{12}{Institute for Cosmic Ray Research, The University of Tokyo, Kashiwa, Chiba 277-8582, Japan}
\altaffiltext{13}{Department of Physics, Graduate School of Science, The University of Tokyo, Bunkyo, Tokyo 113-0033, Japan}
\altaffiltext{14}{Department of Liberal Arts and Sciences, Sapporo Medical University, Chuo, Sapporo 060-8556, Japan.}
\altaffiltext{15}{Institute of Astronomy, The University of Tokyo, Mitaka, Tokyo 181-0015, Japan.}
\altaffiltext{16}{Kavli Institute for the Physics and Mathematics of the Universe, WPI, The University of Tokyo,Kashiwa, Chiba 277-8583, Japan.}
\altaffiltext{17}{Institute for Advanced Research, Nagoya University, Furo-cho, Chikusa-ku, Nagoya 464-8602, Japan.}
\altaffiltext{18}{Koyama Astronomical Observatory, Kyoto-Sangyo University, Kita, Kyoto, 603-8555, Japan.}
\altaffiltext{19}{Graduate School of Science, Nagoya University, Furo-cho, Chikusa-ku, Nagoya 464-8602, Japan.}
\altaffiltext{20}{Institute of Astronomy and Astrophysics, Academia Sinica, Taipei, 10617, Taiwan.}
\altaffiltext{21}{Hiroshima Astrophysical Science Center, Hiroshima University, Higashi-Hiroshima, Hiroshima 739-8526, Japan.}
\altaffiltext{$\star$}{E-mail: yk.matsuoka@nao.ac.jp.}
\altaffiltext{$\dagger$}{Based on data collected at the Subaru Telescope, which is operated by the National Astronomical Observatory of Japan.}
\altaffiltext{$\ddagger$}{Based on observations made with the Gran Telescopio Canarias (GTC), installed at the Spanish Observatorio del Roque de los Muchachos 
of the Instituto de Astrof\'{i}sica de Canarias, on the island of La Palma.}

\begin{abstract}
We report the discovery of 15 quasars and bright galaxies at $5.7 < z < 6.9$.
This is the initial result from the Subaru High-$z$ Exploration of Low-Luminosity Quasars (SHELLQs) project, which 
exploits the exquisite multiband imaging data produced by the Subaru Hyper Suprime-Cam (HSC) Strategic Program survey.
The candidate selection is performed by combining several photometric approaches including a Bayesian probabilistic algorithm to reject stars and dwarfs.
The spectroscopic identification was carried out with the Gran Telescopio Canarias and the Subaru Telescope for the first 80 deg$^2$ of the survey footprint.
The success rate of our photometric selection is quite high, approaching 100 \% at the brighter magnitudes ($z_{\rm AB} < 23.5$ mag).
Our selection also recovered all the known high-$z$ quasars on the HSC images.
Among the 15 discovered objects, six are likely quasars, while the other six with interstellar absorption lines and in some cases narrow emission lines 
are likely bright Lyman-break galaxies.
The remaining three objects have weak continua and very strong and narrow Ly$\alpha$ lines, which may be excited by ultraviolet light from 
both young stars and quasars.
These results indicate that we are starting to see the steep rise of the luminosity function of $z \ge 6$ galaxies, compared with that of quasars, at
magnitudes fainter than $M_{\rm 1450} \sim -22$ mag or $z_{\rm AB} \sim 24$ mag.
Follow-up studies of the discovered objects as well as further survey observations are ongoing.
\end{abstract}





\section{Introduction \label{sec:intro}}

The era from the birth of the first stars to cosmic reionization is one of the key subjects in astronomy and astrophysics today. 
While the formation of the first stars is observationally out of reach at present, the epoch of reionization is being explored with several different approaches.
The latest measurements of the cosmic microwave background (CMB) by the {\it Planck} space mission suggest a reionization optical depth of $\tau = 0.058 \pm 0.012$,
which implies that the midpoint of reionization occurred between $z = 7.8$ and 8.8 \citep{planck16}.
This value is marginally consistent with the rapid decline of the \ion{H}{1} neutral fraction of the intergalactic medium (IGM) from $z \sim 8$ to  $6$,
inferred from the evolving Ly$\alpha$ luminosity function of galaxies \citep[e.g.,][]{ouchi08, ouchi10, konno14, bouwens15, choudhury15}.
The Universe is in the final phase of reionization at $z \sim 6$, as suggested by the small but non-zero \ion{H}{1} transmission fraction of the IGM
implied by the \citet[][GP hereafter]{gunn65} troughs of luminous high-$z$ quasars \citep{fan06araa}.

The main source of the ultraviolet (UV) photons that caused the reionization of the Universe is still under debate. 
It has been argued that star-forming galaxies observed in deep surveys are not able to produce a sufficient number of photons to sustain reionization 
\citep[e.g.,][]{robertson10, robertson13}, while the revised {\it Planck} results with later reionization than previously thought
may alleviate this problem \citep[e.g.,][]{robertson15}. 
Active galactic nuclei (AGNs) have been studied as a possible additional source of ionizing photons \citep[e.g.,][]{lehnert03,fontanot12,grissom14,
giallongo15, madau15}, but the results are still controversial, largely due to the lack of knowledge about the numbers of faint quasars and AGNs residing in the reionization era.

High-$z$ quasars are also a key population to understanding the formation and evolution of supermassive black holes (SMBHs).
If the assembly of an SMBH is predominantly via gas accretion onto a seed black hole with mass $M_{\rm BH, 0}$, then the time needed to grow to the mass $M_{\rm BH}$ is
\begin{eqnarray}
t & = & t_{\rm Edd} \biggl( \frac{\epsilon}{1-\epsilon}\biggr) \lambda_{\rm Edd}^{-1}\ \rm{ln}\biggl(\frac{M_{\rm BH}}{M_{\rm BH, 0}}\biggr) \nonumber \\
  & = & 0.043\ \rm{ln}\biggl(\frac{M_{\rm BH}}{M_{\rm BH,0}}\biggr)\  {\rm Gyr}, \nonumber
\end{eqnarray}
where $t_{\rm Edd} = 0.44 \mu_e^{-1}$ Gyr is the Eddington timescale, $\mu_e$ is the mean molecular weight per electron, $\epsilon$ is 
the radiative efficiency, and $\lambda_{\rm Edd}$ is the Eddington ratio \citep{shapiro05, madau14}; here we assume that $\lambda_{\rm Edd}$ is constant in time.
We adopt $\mu_e = 1.15$, $\epsilon = 0.1$, and $\lambda_{\rm Edd} = 1.0$ to derive the second line of the equation.
For example, a seed with $M_{\rm BH, 0} = 100 M_\odot$ will take 0.7 Gyr to form a quasar with $M_{\rm BH} = 10^9 M_\odot$.
Because this timescale is comparable to the cosmic time that elapsed between $z = 20$ and $z = 6$,
the SMBH mass function at $z > 6$ conveys critical information about the mass distribution of the seed black holes and the mode of subsequent growth,
including super-Eddington accretion \citep{kawaguchi04}.
Indeed, recent discoveries of the luminous quasars ULAS J1120$+$0641 with $M_{\rm BH} \sim 2 \times 10^9 M_\odot$ at $z = 7.085$ \citep{mortlock11}
and SDSS J0100$+$2802 with $M_{\rm BH} \sim 1 \times 10^{10} M_\odot$ at $z = 6.30$ \citep{wu15} have made a significant impact on 
such models
\citep[e.g.,][]{volonteri12, ferrara14, madau14}.

Furthermore, high-$z$ quasars might be a signpost of galaxies and high density peaks in the dark matter distribution in the early Universe.
The stellar and gaseous properties in and around the host galaxies can be studied in the optical/infrared \citep[e.g.,][]{kashikawa07, goto09, goto12, willott11}
or at sub-mm/radio wavelengths \citep[e.g.,][]{maiolino05, wang07, venemans12, venemans16, wang13, willott13, willott15}, 
giving a unique probe of galaxies in the reionization era.
The chemical enrichment, and thus the preceding star formation history, can be measured with strong metal emission lines arising from ionized gas
around the quasar nuclei \citep[e.g.,][]{jiang07, derosa11, derosa14}.
On the whole, no or little chemical evolution of quasars has been observed from $z \sim 7$ to the local Universe.
A good example of this is ULAS J1120$+$0641, 
whose emission-line and continuum spectrum is strikingly similar to those of the local quasars except for the deep GP trough.

In the last two decades there has been great progress in the quest for high-$z$ quasars.\footnote{Hereafter, ``high-$z$" denotes $z > 5.7$,
where quasars are observed as $i$-band dropouts in the Sloan Digital Sky Survey (SDSS) filter system \citep{fukugita96}.}
The SDSS \citep{york00} provided the first opportunity to search for high-$z$ quasars over wide fields ($> 1000$ deg$^2$),
resulting in several tens of objects published to date \citep{fan00, fan01, fan03, fan04, fan06, jiang08, jiang09, jiang15}.
The Canada--France High-$z$ Quasar Survey \citep[CFHQS;][]{willott05, willott07, willott09, willott10a,willott10} 
explored fainter magnitudes than the SDSS and found a few tens of new quasars, including one very faint object (CFHQS J0216$-$0455; $z_{\rm AB} = 24.4$ mag at $z = 6.01$) 
discovered in the Subaru {\it XMM-Newton} Deep Survey \citep{furusawa08} area. 
However, these optical surveys are not sensitive to redshifts beyond $z = 6.5$, where quasars become almost invisible at
observed wavelengths $\lambda_{\rm obs} < 0.9$ $\mu$m due to strong IGM absorption.
The first quasar discovered at $z > 6.5$ was ULAS J1120$+$0641 mentioned above, which was selected from near-infrared (NIR) data of
the United Kingdom Infrared Telescope (UKIRT) Infrared Deep Sky Survey \citep[UKIDSS;][]{lawrence07}.

With the advent of the Visible and Infrared Survey Telescope for Astronomy (VISTA) Kilo-degree Infrared Galaxy (VIKING) survey, more
$z > 6.5$ quasars have been discovered in recent years \citep{venemans13, venemans15b}.
New optical wide-field surveys such as the Panoramic Survey Telescope \& Rapid Response System 1 
\citep[Pan-STARRS1;][]{kaiser10} 3$\pi$ survey and the Dark Energy Survey \citep{des16} are equipped with a $y$-band filter centered at 9500 -- 10000 \AA, 
and are starting to deliver many more quasars at $6 \le z \le 7$ \citep{banados14, reed15}.
In addition, there are various smaller projects which have succeeded in identifying high-$z$ quasars \citep[e.g.,][]{goto06, carnall15, kashikawa15, kim15}.
In total, the above surveys 
have identified about a hundred high-$z$ quasars published to date.
Most of the quasars are located at $z < 6.5$ and $z_{\rm AB} < 22.5$ mag, while the higher redshifts and fainter magnitudes are still poorly explored.
The known bright high-$z$ quasars must be just the tip of the iceberg predominantly composed of faint quasars and AGNs, which may be a significant
contributor to reionization, and may represent the more typical mode of SMBH growth in the early Universe.

This paper describes our ongoing project, SHELLQs (Subaru High-$z$ Exploration of Low-Luminosity Quasars), which is 
the first 1000 deg$^2$ class survey for high-$z$ quasars with a 8 m class telescope.
The project exploits multiband photometry data produced by the Subaru Hyper Suprime-Cam (HSC) Subaru Strategic Program (SSP) survey.
We present the results of the initial follow-up spectroscopy of photometric candidates, performed in the 2015 Fall and 2016 Spring semesters, which delivered
15 high-$z$ objects including both quasars and bright galaxies.
This paper is organized as follows.
We introduce the Subaru HSC-SSP survey in \S 2.
The details of the photometric candidate selection are presented in \S 3.
The spectroscopic follow-up observations are described in \S 4.
The quasars and galaxies we have discovered are presented and discussed in \S 5.
The summary appears in \S 6.
We adopt the cosmological parameters $H_0$ = 70 km s$^{-1}$ Mpc$^{-1}$, $\Omega_{\rm M}$ = 0.3, and $\Omega_{\rm \Lambda}$ = 0.7.
All magnitudes in the optical and NIR bands are presented in the AB system \citep{oke83}.
Magnitudes refer to point spread function (PSF) magnitudes (see \S 2) 
unless otherwise noted.

\section{The Subaru HSC-SSP Survey \label{sec:hscsurvey}}

The Subaru HSC-SSP survey (M. Takada et al. 2016, in preparation) is a large collaborative project with contributions from researchers in Japan, Taiwan, and Princeton University.
The project started in early 2014, and will include 300 nights until around 2019.
It uses the HSC \citep[][S. Miyazaki et al. 2016, in preparation]{miyazaki12}, a wide-field camera newly installed on the Subaru 8.2 m telescope on the summit of Maunakea.
HSC is equipped with 116 2K $\times$ 4K Hamamatsu fully depleted CCDs, of which 104 CCDs are used to obtain science data.
The pixel scale is 0\arcsec.17.
The camera has a nearly circular field of view of 1$^\circ$.5 diameter, which enables it to image 1.77 deg$^2$ of the sky in a single shot.
Five broad-band filters ($g$, $r$, $i$, $z$, and $y$) and several narrow-band filters are currently available.

The HSC-SSP survey has three layers with different combinations of area and depth.
The Wide layer aims to observe 1400 deg$^2$ mostly along the celestial equator through the five broad-band filters.
The present paper is based on this Wide-layer data.
The total exposure times range from 10 minutes in the $g$-, $r$-bands to 20 minutes in the $i$-, $z$-, and $y$-bands, divided into individual exposures of $\sim$3 minutes each.
The target 5$\sigma$ limiting magnitudes are ($g$, $r$, $i$, $z$, $y$) = (26.5, 26.1, 25.9, 25.1, 24.4) mag measured in 2\arcsec.0 apertures.
The Deep and Ultra-Deep layers observe 27 and 3.5 deg$^2$, respectively, within and around popular deep survey fields.
Five broad-band filters and four narrow-band filters are used, aiming to reach the 5$\sigma$ limiting depth of $r = 27.1$ mag (Deep) 
or $r = 27.7$ mag (Ultra-Deep).

The SHELLQs project exploits the exquisite HSC survey data to search for low-luminosity quasars at high redshift.
Assuming the quasar luminosity function at $z \ge 6$ presented by \citet{willott10}, 
the expected numbers of newly identified quasars in the Wide layer are $\sim$500 with $z_{\rm AB} < 24.5$ mag at $z \sim 6$ and 
$\sim$100 with $y_{\rm AB} < 24.0$ mag at $z \sim 7$.
The former magnitude limit corresponds to $M_{1450} < -22$ mag for a typical quasar spectral energy distribution (SED), thus allowing us to explore
$\sim$2 mag lower luminosity than any previous wide-field survey at $z \sim 6$ \citep[e.g.,][]{jiang09}.
Our filter set is sensitive to quasars with redshifts up to $z \sim 7.4$ (i.e., beyond the current quasar redshift record). 
However, the detection capability sharply drops at $z > 7$ where the GP trough comes into the $y$-band, hence the survey is limited to intrinsically
very luminous objects at those redshifts.
The deep optical data produced by the HSC survey will also provide opportunities to explore even higher redshifts when combined with wide and deep NIR surveys.

This paper describes the results from the early HSC survey data taken before 2015 August.
It covers roughly 80 deg$^2$ in the five broad bands in the Wide layer, with a median seeing of 0\arcsec.6 -- 0\arcsec.8.
Data reduction was performed with the dedicated pipeline {\tt hscPipe} (version 3.8.5; J. Bosch et al. 2016, in preparation) derived from
the Large Synoptic Survey Telescope (LSST) software pipeline 
\citep{ivezic08, axelrod10, juric15}, 
for all the standard procedures
including bias subtraction, flat fielding with dome flats, stacking, astrometric and photometric calibrations, and source detection and measurements.
The astrometric and photometric calibrations are tied to the Pan-STARRS1 system 
 \citep{schlafly12,tonry12,magnier13}.
We utilize forced photometry, which allows for flux measurements in all the five bands with a consistent aperture defined in a reference band.
The reference band is $i$ by default and is switched to $z$ ($y$) for extremely red sources with no detection in the $i$ ($z$) and bluer bands.
We use the PSF magnitude ($m_{\rm PSF}$, or simply $m$) and the cModel magnitude ($m_{\rm cModel}$),
which are measured by fitting the PSF models and two-component, PSF-convolved galaxy models to the source profile, respectively \citep{abazajian04}.
We measure fluxes and colors of sources with $m_{\rm PSF}$, while the source extendedness is evaluated with $m_{\rm PSF} - m_{\rm cModel}$.
All the magnitudes are corrected for Galactic extinction \citep{schlegel98}.



We performed a rough assessment of the completeness limits achieved in the early Wide survey as follows.
In each subregion of the sky with an approximate size of 12\arcmin $\times$ 12\arcmin, 
we select every source whose processing flags indicate clean photometry, and measure the number counts $N(m)$.
It typically follows a straight line in the $m$ -- log$\bigl[ N(m)\bigr]$ plane at bright magnitudes, then peaks at $m = m_{\rm peak}$.
We fit a straight line $\widetilde{N}(m)$ to the $m$ -- log$\bigl[ N(m)\bigr]$ relation at $m_{\rm peak} - 5 < m < m_{\rm peak}  - 1$,
and define the 50 \% completeness magnitude at the point where $N(m)$ falls to half  of $\widetilde{N}(m)$.
The resultant completeness magnitudes averaged over all the fields are (26.5, 26.3, 26.4, 25.5, 24.7) mag in the ($g$, $r$, $i$, $z$, $y$)-band, respectively, 
with a typical field-to-field variation of 0.3 mag.
The spatial pattern of the above estimates agrees with that of the seeing, in such a way that poor seeing is accompanied by worse-than-average depth.
These completeness magnitudes are roughly 0.3 mag fainter than the target limiting magnitudes of the survey mentioned above, which is at least partly because of 
the differences in the adopted magnitudes (PSF magnitudes vs. 2\arcsec.0-aperture magnitudes) and in the definitions of the depths (50\% completeness
vs. 5$\sigma$ detection).

Our project also benefits from archival NIR data from UKIDSS and VIKING.
The UKIDSS is a multi-tiered imaging survey using WFCAM, a wide-field camera mounted on the UKIRT 3.8 m telescope \citep{lawrence07}.
The widest Large Area Survey covers most of the HSC survey footprint, with target 5$\sigma$ limiting magnitudes of ($Y$, $J$, $H$, $K$) = 
(20.9, 20.4, 20.0, 20.1) mag measured in 2\arcsec.0 apertures.
The VIKING is one of the public surveys of the European Southern Observatory with the VISTA 4.1 m telescope.
This project aims to observe 1500 deg$^2$ of the sky, with target 5$\sigma$ limiting magnitudes of ($Z$, $Y$, $J$, $H$, $K$) = 
(23.1, 22.3, 22.1, 21.5, 21.2) mag measured in 2\arcsec.0 apertures.
Roughly half of the HSC survey footprint will be covered by the VIKING at its completion.
The present work uses the data from the UKIDSS data release 10 and the VIKING data release 4.

\section{Photometric Candidate Selection \label{sec:selection}}

\begin{figure*}
\epsscale{0.9}
\plotone{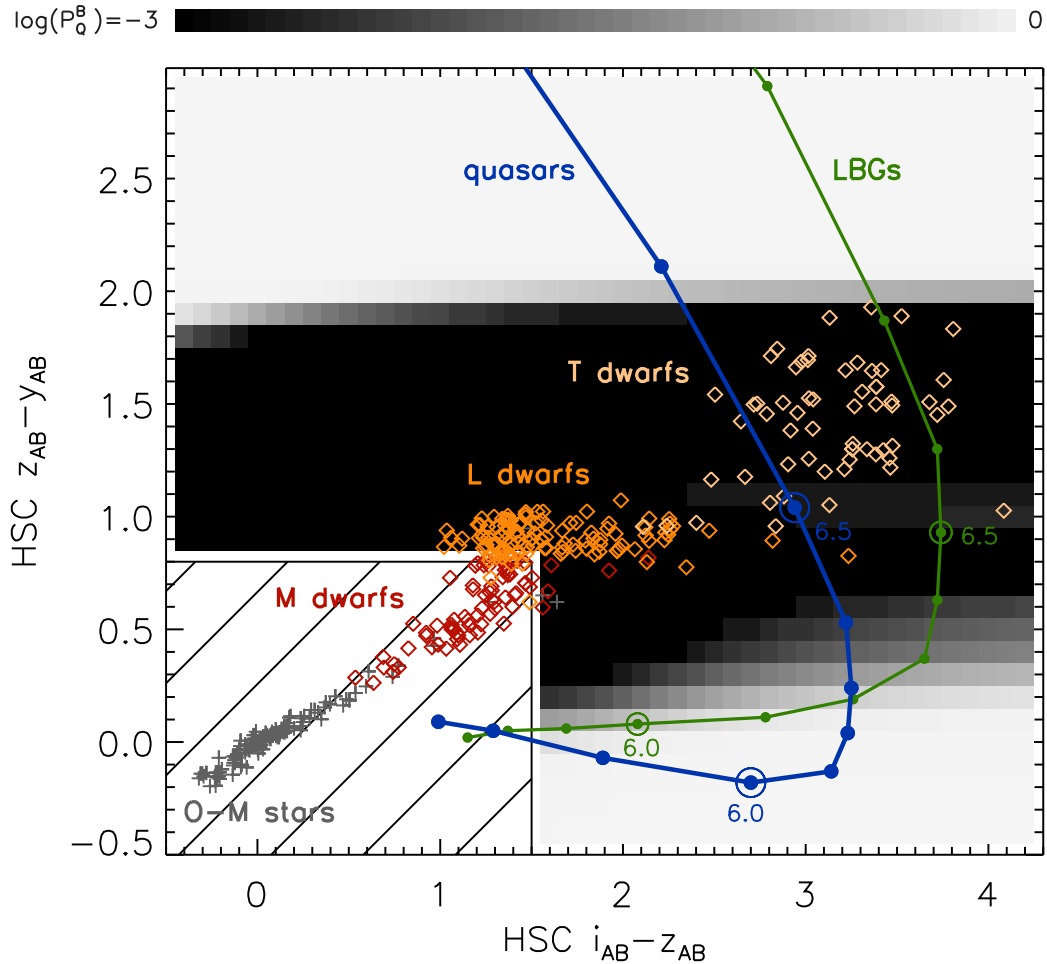}
\caption{The $i - z$ and $z - y$ colors of high-$z$ quasars (blue line) and galaxies (green line), as well as Galactic stars and brown dwarfs (crosses and diamonds).
The SED models for quasars and brown dwarfs are described in \S \ref{subsec:mainsample}.
The galaxy colors are calculated with the spectral templates taken from \citet{gonzalez12}, while the colors of O -- M stars are computed with the \citet{pickles98} library.
The dots along the blue and green lines represent redshifts in steps of 0.1, with $z = 6.0$ and $6.5$ marked by the large open circles.
The hatched area in the lower left indicates the color space excluded from the HSC-SSP database query in the first step of our quasar selection (\S \ref{subsec:flow}).
For example, the grayscale represents the Bayesian quasar probabilities $P_{\rm Q}^{\rm B}$ 
(Equations \ref{eq:bayes1} and \ref{eq:bayes2}; the color bar is found at the top)
over this plane.
Note that the $P_{\rm Q}^{\rm B}$ distribution changes in response to the source and limiting magnitudes;
here we assume a source magnitude of $z_{\rm AB} = 24.0$ mag and the 5$\sigma$ limiting magnitudes of ($i_{\rm AB}$, $z_{\rm AB}$, $y_{\rm AB}$) = (26.5, 25.5, 25.0) mag.
Galaxy models are not included in the Bayesian algorithm at present.
\label{fig:colordiagram}}
\end{figure*}

\begin{figure*}
\epsscale{1.0}
\plotone{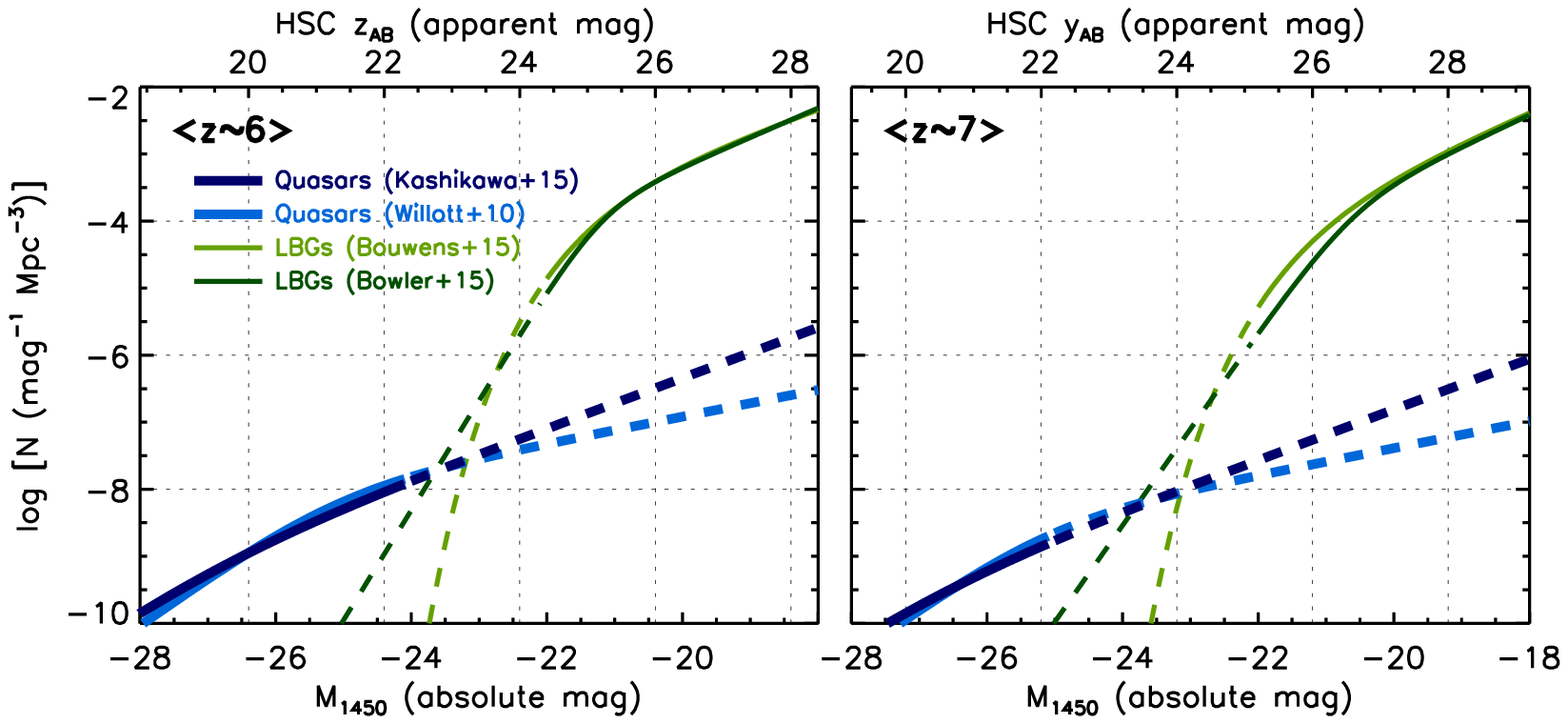}
\caption{
Luminosity functions of quasars and LBGs at $z \sim 6$ (left) and $z \sim 7$ (right), taken from
\citet[][dark blue; their case 2]{kashikawa15}, \citet[][light blue]{willott10}, \citet[][light green]{bouwens15}, and \citet[][dark green]{bowler15}.
The corresponding apparent magnitudes for quasars in the HSC $z$- or $y$-band are indicated on the upper axis.
Although the luminosity functions are poorly constrained in the ranges marked by the dashed lines, LBGs likely outnumber quasars at apparent 
magnitudes fainter than $\sim$24 mag.
\label{fig:numbercounts}}
\end{figure*}

High-$z$ quasars are characterized by extremely red optical colors caused by strong IGM absorption blueward of Ly$\alpha$. 
This is demonstrated in Figure \ref{fig:colordiagram}, which presents $i - z$ and $z - y$ colors of model quasars and other populations in the HSC passband system.
There are three major sources of astrophysical contamination to the photometric selection of quasars. 
The first is Galactic brown dwarfs, which have been the most serious contaminants in past surveys because of their very red colors 
and point-like appearance. 
The second is red galaxies at $z \sim 1$, whose 4000 \AA\ break leads to red $i - z$ colors, but we expect that
the excellent image quality of the HSC will help identify those low-$z$ galaxies morphologically.
The third is faint Lyman-break galaxies (LBGs) at $z \ge 6$, which are also affected by IGM absorption.  
Figure \ref{fig:numbercounts} displays the luminosity functions of high-$z$ quasars \citep{willott10, kashikawa15} and LBGs \citep{bouwens15, bowler15}.
We assume the galaxy UV spectral slope of $\beta = -2$ \citep[e.g.,][]{stanway05} to convert the UV magnitudes in the literature to $M_{1450}$.
Although the luminosity functions are still poorly constrained at $M_{1450} > -24$ mag for quasars and at $< -22$ mag for LBGs, it is likely that they intersect 
at an apparent magnitude of $\sim$24 mag, with LBGs outnumbering quasars at fainter magnitudes.
This is why previous surveys at brighter magnitudes did not suffer from severe LBG contamination. 
The LBG contamination in our project could be significant, particularly if the LBG luminosity function has a double power-law form instead of the Schechter function
with its exponential cut-off at the bright end \citep{bowler14, bowler15}.

There are several methods to extract quasar candidates from multiband imaging data.
Our sample of candidates is selected based on a Bayesian probabilistic approach, as detailed in \S \ref{subsec:mainsample}.
It computes the posterior probability for each source being a high-$z$ quasar rather than a red star or dwarf, based on photometry in all the available bands 
as well as SED and surface density models of the populations under consideration.
We use photometry in the HSC optical bands plus the NIR bands ($Y$, $J$, $H$, $K$) where available from UKIDSS or VIKING.
While UKIDSS is too shallow for all but the brightest HSC sources, 
the VIKING data are useful to identify and remove brown dwarfs, because the mean colors of L--T dwarfs ($z_{\rm AB} - J_{\rm AB} \sim 2 - 4$ mag) match
the relative depths of the HSC survey and VIKING.
The entire flow of the candidate selection from the HSC-SSP database to the final spectroscopic targets is described in \S \ref{subsec:flow}.
Our selection procedure has efficiently removed contaminants, recovered all the known quasars imaged by the HSC so far (\S \ref{subsec:recovery}),
and discovered a number of new objects as described in the following sections. 

\subsection{Bayesian algorithm \label{subsec:mainsample}}

Our quasar candidates are selected with a Bayesian probabilistic algorithm, developed following \citet{mortlock12}.
Here we assume that galaxies and relatively blue stars with O to early-M spectral types have been removed in advance with color and extendedness cuts 
(see \S \ref{subsec:flow}). 
Although this may not be the case for LBGs at $z \ga 6$, we do not model this population at present because (i) they are hard to distinguish from high-$z$ quasars
by colors alone and (ii) their surface density at $< 24$ mag is still poorly understood (see Figures \ref{fig:colordiagram} and \ref{fig:numbercounts}).

For a detected source with the observed quantities $\vec{d}$, the Bayesian probability $P_{\rm Q}^{\rm B}$ 
of being a quasar is given by:
\begin{eqnarray}
P_{\rm Q}^{\rm B} (\vec{d})\ = \ \frac{W_{\rm Q} (\vec{d})}{W_{\rm Q} (\vec{d}) + W_{\rm D} (\vec{d})} 
\label{eq:bayes1}
\end{eqnarray}
and
\begin{eqnarray}
W_{\rm Q/D} (\vec{d})\ = \ \int S (\vec{p})\ Pr ( {\rm det} | \vec{p} )\ Pr (\vec{d} | \vec{p})\ d\vec{p}  
\label{eq:bayes2}
\end{eqnarray}
where the subscripts Q and D denote a quasar and a brown dwarf, respectively.
The vector $\vec{d}$ represents the magnitudes in all the available bands in the present case, while $\vec{p}$ represents the intrinsic source properties (i.e., 
luminosity and redshift for a quasar and luminosity and spectral type for a brown dwarf).
The functions $S (\vec{p})$, $Pr ( {\rm det} | \vec{p} )$, and $Pr (\vec{d} | \vec{p})$ represent the surface number density, the probability that the source is detected (``det"), 
and the probability that the source has the observed quantities $\vec{d}$, respectively, each as a function of $\vec{p}$.

We compute $S (\vec{p})$ with the quasar luminosity function of \citet{willott10} and the Galactic brown dwarf model of \citet{caballero08}.
The former is well determined at $M_{1450} < -24$ mag, while we extrapolate it to $M_{1450} = -20$ mag as shown in Figure \ref{fig:numbercounts} 
in order to match the HSC observations.
The brown dwarf model takes into account the spatial density distributions and luminosities of late-M, L, and T dwarfs, and allows one to compute 
number counts for each spectral type at a given Galactic coordinate.
At our quasar selection limit of $z_{\rm AB} = 24.5$ mag (see below), L--T dwarfs within $\sim$1 kpc of the Sun are bright enough to enter our sample.
The validity of these quasar and brown dwarf models will be evaluated with the results of our and other surveys in future work.

Since the HSC-SSP survey depth has not been fully analyzed yet (\S 2), we arbitrarily set  $Pr ( {\rm det} | \vec{p} ) = 1$ 
for $z_{\rm AB} < 26.0$ or $y_{\rm AB} < 25.0$ mag and 0 otherwise.
The SED models required for $Pr (\vec{d} | \vec{p})$ are created as follows.
The quasar model spectrum at $z = 0$ is first created by stacking the SDSS spectra of 340 bright quasars at $z \simeq 3$, where the quasar selection is 
fairly complete \citep{richards02, willott05}, after correcting for the foreground IGM absorption.
The IGM \ion{H}{1} opacity data are taken from \citet{songaila04}.
This spectrum is then placed at various redshifts with the appropriate IGM absorption taken into account, and convolved with the filter transmission functions
to compute colors.
Because the model spectra redshifted to $z > 6$ do not extend beyond the $J$-band, we take the $J - H$ and $H - K$ colors from \citet{hewett06}.
The dwarf colors are computed with a set of observed spectra compiled in the SpeX prism library\footnote{
This research has benefited from the SpeX Prism Spectral Libraries, maintained by Adam Burgasser at http://pono.ucsd.edu/$\sim$adam/browndwarfs/spexprism.}
and in the CGS4 library.\footnote{The L and T dwarf archive is maintained by Sandy Leggett at http://staff.gemini.edu/$\sim$sleggett/LTdata.html.}
Because of the discrete sampling of the brown-dwarf templates grouped into individual spectral types, the integration in Equation \ref{eq:bayes2} is treated 
as a summation for spectral types.
Finally, the flux errors are taken from the outputs of the HSC image processing pipeline, and are assumed to follow a Gaussian probability density distribution
in fluxes.

\subsection{Selection flow \label{subsec:flow}}

The present work is based on the HSC-SSP Wide-layer data included in the S15A internal data release, which happened in 2015 September.
Forced photometry (see \S 2) on the stacked images is used.
We first query the HSC-SSP database for non-blended\footnote{
{In particular, we query sources that are isolated or deblended from parent blended sources, and we reject those parents.
This corresponds to \tt deblend.nchild = 0} in the database language.} sources meeting the following criteria:
\begin{eqnarray}
(z_{\rm AB} < 24.5\ {\rm and}\ \sigma_z < 0.155\ {\rm and}\ i_{\rm AB} - z_{\rm AB} > 1.5 \nonumber \\
 {\rm and}\ z_{\rm AB} - z_{\rm cModel, AB} < 0.3 )
\label{eq:query1}
 \end{eqnarray}
 or
 \begin{eqnarray}
(y_{\rm AB} < 24.0\ {\rm and}\ \sigma_y < 0.155\ {\rm and}\ z_{\rm AB} - y_{\rm AB} > 0.8 \nonumber \\
 {\rm and}\ y_{\rm AB} - y_{\rm cModel, AB} < 0.3 ) 
\label{eq:query2}
\end{eqnarray}
and without any critical quality flags assigned.\footnote{
Specifically, we require that the source is (i) not close to an edge of the processed image frame ({\tt flags.pixel.edge} = {\tt False}), 
(ii) not in a bad CCD region (i.e., bad pixels or vignetted area; {\tt flags.bad.center} = {\tt False}), 
(iii) not saturated ({\tt flags.pixel.saturated.center} = {\tt False}),
and (iv) not affected by cosmic rays ({\tt flags.pixel.cr.center} = {\tt False}) in the $i$-, $z$-, and $y$-bands.
}
Throughout this paper $i_{\rm AB}$, $z_{\rm AB}$, and $y_{\rm AB}$ refer to PSF magnitudes.
The conditions of Equation \ref{eq:query1} select $i$-band dropouts at $z \sim 6$, while those of Equation \ref{eq:query2} select $z$-band dropouts at $z \sim 7$.
The color cuts are used to exclude relatively blue stars with O to early-M spectral types (see Figure \ref{fig:colordiagram}),
while the difference between the PSF and cModel magnitudes is used to exclude extended sources.
After the database query, we further remove low-$z$ interlopers with more than $3\sigma$ detection in the $g$- or $r$-band.

Because we are looking for rare faint objects detected in only one or two bands, we are sensitive to false dropout sources
of both astrophysical and non-astrophysical origins.
However, we can take advantage of the fact that 
in the HSC-SSP Wide survey, each patch of the sky is visited and imaged several times with different dithering positions.
Glitches in the data may become apparent upon comparing the individual per-visit exposures.
For every candidate, we retrieve and perform photometry on all the per-visit images with Source Extractor, version 2.8.6 \citep{bertin96}, 
in the double-image mode, with the stacked image as the detection reference.
If any of the per-visit photometric measurements deviate by more than three times the measurement error from the stacked photometry, the candidate is eliminated.
This procedure is performed in the band in which the source photometry has the highest signal-to-noise ratio (S/N; typically the $z$ -and for $i$ dropouts and the $y$-band for $z$ dropouts).
We also reject candidates with profiles that are too compact, diffuse, or elliptical to be celestial point sources with the Source Extractor measurements
on the stacked images.
The eliminated sources are mostly cosmic rays, moving or transient sources, and image artifacts.

The candidates selected above are matched to the UKIDSS and VIKING catalogs within 1\arcsec.0 in the overlapping survey area.
They are then processed through the Bayesian probabilistic algorithm, and
those with the quasar probability $P_{\rm Q}^{\rm B} > 0.1$ are added to the sample of candidates.
The rather low value of the threshold $P_{\rm Q}^{\rm B} = 0.1$ was chosen to ensure that we would not throw away any possible candidates.
We found that the actual $P_{\rm Q}^{\rm B}$ distribution is bimodal, with only a small fraction falling in  $0.1 < P_{\rm Q}^{\rm B} < 0.9$,
so our results are insensitive to the exact value of this cut.

Finally, we inspect images of all the candidates by eye and reject additional problematic objects.
HSC stacked and per-visit images are used for this purpose. 
The sources rejected at this stage include those close to very bright stars, cosmic rays, and moving objects overlooked in the above automatic procedure.

In the present survey area covering 80 deg$^2$, we had roughly 50,000 red and point-like sources meeting the database query conditions (Equations \ref{eq:query1} and \ref{eq:query2}) 
and undetected in the $g$- and $r$-bands.
The vast majority of them ($\sim$97 \%) were eliminated by checking the per-visit photometry and source morphology as described above.
From $\sim$2000 remaining candidates, the Bayesian algorithm selected 117 candidates with $P_{\rm Q}^{\rm B} > 0.1$.
In this initial work, we further excluded candidates with bluer colors ($1.5 < i_{\rm AB} - z_{\rm AB} < 2.0$ and $z_{\rm AB} - y_{\rm AB} < 0.8$; see Figure \ref{fig:colordiagram}),
fainter magnitudes ($z_{\rm AB} >$ 24.3 mag), or with detection only in the $z$-band; we obtained 38 final candidates after visual image inspection.

We are also developing more classical methods of quasar color selection, such as simple color cuts and SED fitting with no Bayesian prior, in order to understand completeness
and any possible bias in the Bayesian algorithm.
A comparison of these selection techniques will be presented in a future paper.
We found that the present sample of 38 candidates passed our current color cut and SED fitting selection criteria as well.

\subsection{Recovery of known quasars \label{subsec:recovery}}

\begin{figure}
\epsscale{1.1}
\plotone{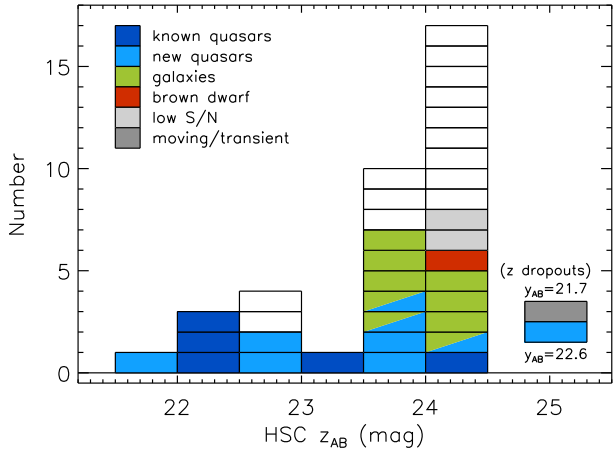}
\caption{Magnitude histogram of the 38 candidates.
The five dark blue cells represent the high-$z$ quasars known prior to our survey.
The remaining cells are color-coded according to the results of our follow-up spectroscopy (see \S 4) as follows.
Light blue: high-$z$ quasars, green: high-$z$ galaxies, red: brown dwarf, dark gray: moving object or transient event, light gray: ambiguous 
nature due to the low spectral S/N, and
white: awaiting spectroscopy.
The two cells in lower right represent $z$-band dropouts.
The success rate of our photometric quasar selection is quite high, approaching 100 \% at the brighter magnitudes ($z_{\rm AB} < 23.5$ mag).
Most of the contaminants at fainter magnitudes are high-$z$ galaxies.
\label{fig:maghistogram}}
\end{figure}

Figure \ref{fig:maghistogram} presents the magnitude histogram of the sample selected above.
We found 36 $i$-band dropouts interpreted as $z \sim 6$ quasar candidates, and two $z$-band dropouts as $z \sim 7$ quasar candidates.
The magnitudes of the former objects range from $z_{\rm AB} = 21.8$ mag to the limiting magnitude of our selection, $z_{\rm AB} = 24.5$ mag, 
while the latter objects are fairly bright in the $y$-band, 21.7 and 22.6 mag.

Five high-$z$ quasars in the present survey area were identified prior to our project;
they are summarized in Table \ref{tab:knownquasars}.
We found that they all successfully pass our selection criteria and end up in the final quasar candidates, as marked by the dark blue cells in Figure \ref{fig:maghistogram}.
The Bayesian quasar probability is $P_{\rm Q}^{\rm B} = 1.000$ in all cases. 
The figure implies that the success rate of our quasar selection is quite high at the brightest magnitudes; for example, three of the four candidates 
at $z < 22.5$ mag were indeed previously discovered high-$z$ quasars.
The five known quasars include CFHQS J0216$-$0455 with $z_{\rm AB} = 24.22$ mag, which, as we discussed in \S 1, was the faintest high-$z$ quasar known before this work.

Thus our candidate selection works quite efficiently at the brighter magnitudes.
We now start to explore the remainder of the sample with new spectroscopic observations, as described in the next section.

\begin{table*}
\begin{center}
\caption{Known high-$z$ quasars in the HSC-SSP S15A footprint.\label{tab:knownquasars}}
\begin{tabular}{cccccl}
\tableline\tableline
R.A. & Decl. &  $i_{\rm AB}$ (mag) & $z_{\rm AB}$ (mag) & $y_{\rm AB}$ (mag) & Comment\\
\tableline
02:10:13.19 & $-$04:56:20.7 & 26.96 $\pm$ 0.42 & 22.34 $\pm$ 0.01 & 22.39 $\pm$ 0.04 & $z = 6.44$; ref (1)\\
02:16:27.79 & $-$04:55:34.1 & 27.61 $\pm$ 0.69 & 24.20 $\pm$ 0.07 & 25.81 $\pm$ 0.68 & $z = 6.01$; ref (2)\\
02:27:43.30 & $-$06:05:30.3 & 25.49 $\pm$ 0.12 & 22.05 $\pm$ 0.01 & 22.03 $\pm$ 0.03 & $z = 6.20$; ref (2) \\
22:19:17.22 & $+$01:02:49.0 & 27.78 $\pm$ 0.80 & 23.43 $\pm$ 0.04 & 23.29 $\pm$ 0.07 & $z = 6.16$; ref (3)\\
22:28:43.52 & $+$01:10:32.0 & 24.01 $\pm$ 0.02 & 22.40 $\pm$ 0.01 & 22.47 $\pm$ 0.02 & $z = 5.95$; ref (4)\\
 \tableline
\end{tabular}
\tablecomments{The coordinates (J2000.0) and magnitudes were measured with the HSC data. 
All the objects have the Bayesian quasar probability $P_{\rm Q}^{\rm B} = 1.000$.
References: (1) \citet{willott10a}, (2) \citet{willott09}, (3) \citet{kashikawa15}, (4) \citet{zeimann11}.}
\end{center}
\end{table*}

\section{Spectroscopy \label{sec:spectroscopy}}

We carried out spectroscopic follow-up observations of the quasar candidates in the 2015 Fall and 2016 Spring semesters.
One of the candidates was observed with the Optical System for Imaging and low-intermediate-Resolution Integrated Spectroscopy \citep[OSIRIS;][]{cepa00}
mounted on the Gran Telescopio Canarias (GTC) in 2015 September, which led to the first discovery of a high-$z$ quasar from our project.
We also observed 19 candidates with the Faint Object Camera and Spectrograph \citep[FOCAS;][]{kashikawa02}  mounted on Subaru in 2015 November and December,
and identified 14 more quasars and galaxies.
We further obtained additional exposures for a few of the above objects in 2016 February.
The journal of these discovery observations is presented in Table \ref{tab:obsjournal}.
The details of the observations are described in the following sections.

\begin{deluxetable}{lllll}
\tablewidth{0pt}
\tablecaption{Journal of discovery observations.\label{tab:obsjournal}}
\tablehead{
\colhead{Target}  & \colhead{Date}  & \colhead{} & \colhead{} & \colhead{}  }
\startdata
\cutinhead{GTC/OSIRIS; 2015 Sep}
                                     & Sep 13 & Sep 14 & & \\
HSC J2216$-$0016 & 90 min & 55 min  & & \\
\cutinhead{Subaru/FOCAS; 2015 Nov and Dec}
                                       & Nov 3    & Nov 4     & Dec 6     & Dec 7  \\
HSC J0210$-$0523  & 23 min & \nodata & 60 min & \nodata \\
HSC J0210$-$0559  & 80 min & \nodata & 45 min & \nodata \\
HSC J0215$-$0555  & \nodata & 40 min$^{\rm s,t}$ & 120 min & 60 min \\
HSC J0219$-$0416   & 80 min & \nodata & \nodata & 60 min \\
HSC J0848$+$0045  & 80 min$^{\rm t}$ & 30 min$^{\rm s}$ & 30 min & \nodata \\
HSC J0850$+$0012  & \nodata & \nodata & 30 min & \nodata \\
HSC J0857$+$0142  & \nodata & \nodata & \nodata & 100 min \\
HSC J0859$+$0022  & \nodata & \nodata & 30 min & \nodata \\
HSC J1152$+$0055  & \nodata & \nodata & 15 min & \nodata \\
HSC J1202$-$0057   & \nodata & \nodata & \nodata & 60 min \\
HSC J1205$-$0000   & \nodata & \nodata & 100 min$^{\rm s}$ & \nodata \\
HSC J1207$-$0005  & \nodata & \nodata & \nodata & 40 min \\
HSC J2216$-$0016  & \nodata & \nodata & \nodata & 30 min \\
HSC J2228$+$0128  & 80 min & \nodata & \nodata & \nodata \\
HSC J2232$+$0012  & 60 min & \nodata & \nodata & 50 min$^{\rm e}$ \\
HSC J2236$+$0032  & 25 min$^{\rm s}$ & 80 min & \nodata & \nodata \\
\cutinhead{Subaru/FOCAS; 2016 Feb}
                                      & Feb 13    & Feb 14     & Feb 15     & Feb 16  \\
HSC J0210$-$0523   & \nodata & \nodata & 30 min$^{\rm e}$ & 40 min \\
HSC J1202$-$0057   & 50 min & \nodata & 60 min & \nodata \\
HSC J1205$-$0000   & \nodata & \nodata & 80 min & 100 min$^{\rm s}$ \\
HSC J1207$-$0005   & \nodata & 60 min & \nodata & \nodata \\
             \enddata
\tablecomments{Observing conditions: $^{\rm s}$poor seeing (1\arcsec.0 $\sim$ 2\arcsec.0), $^{\rm t}$low transparency, $^{\rm e}$low elevation ($\sim$30$^\circ$).}
\end{deluxetable}

\subsection{GTC/OSIRIS}

GTC is a 10.4 m telescope located at the Observatorio del Roque de los Muchachos in La Palma, Spain.
Our program (GTC19-15B; Iwasawa et al.) was awarded 14.4 hr in the 2015B semester.
We used OSIRIS with the R2500I grism and 1\arcsec.0 wide long slit, which provides spectral coverage from $\lambda_{\rm obs}$ = 0.74 to 1.0\ $\mu$m 
with a resolution $R \sim 1500$.
The observations were carried out in queue mode on dark nights with excellent weather conditions and the seeing 0\arcsec.7 -- 1\arcsec.0.

The data were reduced using the Image Reduction and Analysis Facility (IRAF\footnote{IRAF is distributed by the National 
Optical Astronomy Observatory, which is operated by the Association of Universities for Research in Astronomy (AURA) under a cooperative agreement 
with the National Science Foundation.}).
Bias correction, flat fielding with dome flats, sky subtraction, and 1d extraction were performed in the standard way.
The wavelength was calibrated with reference to sky emission lines.
The flux calibration was tied to the white dwarf standard stars, Ross 640 and G191-B2B, observed on the same nights.
We corrected for slit losses by scaling the spectra to match the HSC $z$-band magnitudes.

One of the quasar candidates was observed in this GTC run and identified to be a quasar at $z = 6.10$.
Its spectrum is presented in \S 5.
In addition, we observed five candidates that were selected from the older (S14B) version of the HSC-SSP data release.
These five objects are no longer quasar candidates with the revised photometry in our fiducial (S15A) release, and hence are not included in the sample of
38 candidates described above.
Indeed, their OSIRIS spectra show the relatively smooth red continua characteristics of brown dwarfs, which is also consistent with the latest HSC magnitudes.
The detailed analysis of these objects is still underway and will be reported in a future paper.

\subsection{Subaru/FOCAS}

Our program was awarded four nights in the S15B semester (S15B-070; Matsuoka et al.) and five nights in the S16A semester  
(S16A-076; Matsuoka et al.) with the Subaru 8.2 m telescope.
We used FOCAS in the multiobject spectrograph mode with the VPH900 grism and SO58 order-sorting filter.
The widths of the slitlets were set to 1\arcsec.0.
This configuration provides spectral coverage from $\lambda_{\rm obs}$ = 0.75 to 1.05\ $\mu$m with a resolution $R \sim 1200$.
The observations were carried out in gray nights in 2015 November, December, and 2016 February.
A few of these nights were occasionally affected by cirrus and poor seeing (1\arcsec.0 $\sim$ 2\arcsec.0), while the weather was fairly good
with seeing 0\arcsec.6 -- 0\arcsec.8 in the rest of the observations.

The data were reduced with IRAF using the dedicated FOCASRED package.
Bias correction, flat fielding with dome flats, sky subtraction, and 1d extraction were performed in the standard way.
The wavelength was calibrated with reference to the sky emission lines.
The flux calibration was tied to the white dwarf standard star Feige 110 observed on the same nights as the targets.
We corrected for slit losses by scaling the spectra to match the HSC magnitudes in the $z$-band for the $i$-band dropouts and $y$-band for the 
one $z$-band dropout we observed.

We observed 19 targets from the sample of candidates, including the quasar identified with GTC/OSIRIS, and identified 14 new high-$z$ quasars and galaxies,
as well as a brown dwarf.
Their final spectra are presented in \S 5.
One of the $z$-band dropout targets was not found in the HSC position at the time of the spectroscopy, so is most likely a moving object or a transient event
caught by the HSC $y$-band observations.
All the $y$-band exposures of this object were taken in a single day, and our inspection of per-visit images did not detect significant day-scale
motion or flux variation.
The spectral S/N of the remaining two targets was too low to judge their nature at this moment.

\section{Discovery of high-$z$ quasars and galaxies \label{sec:results}}

\begin{figure*}
\epsscale{1.0}
\plotone{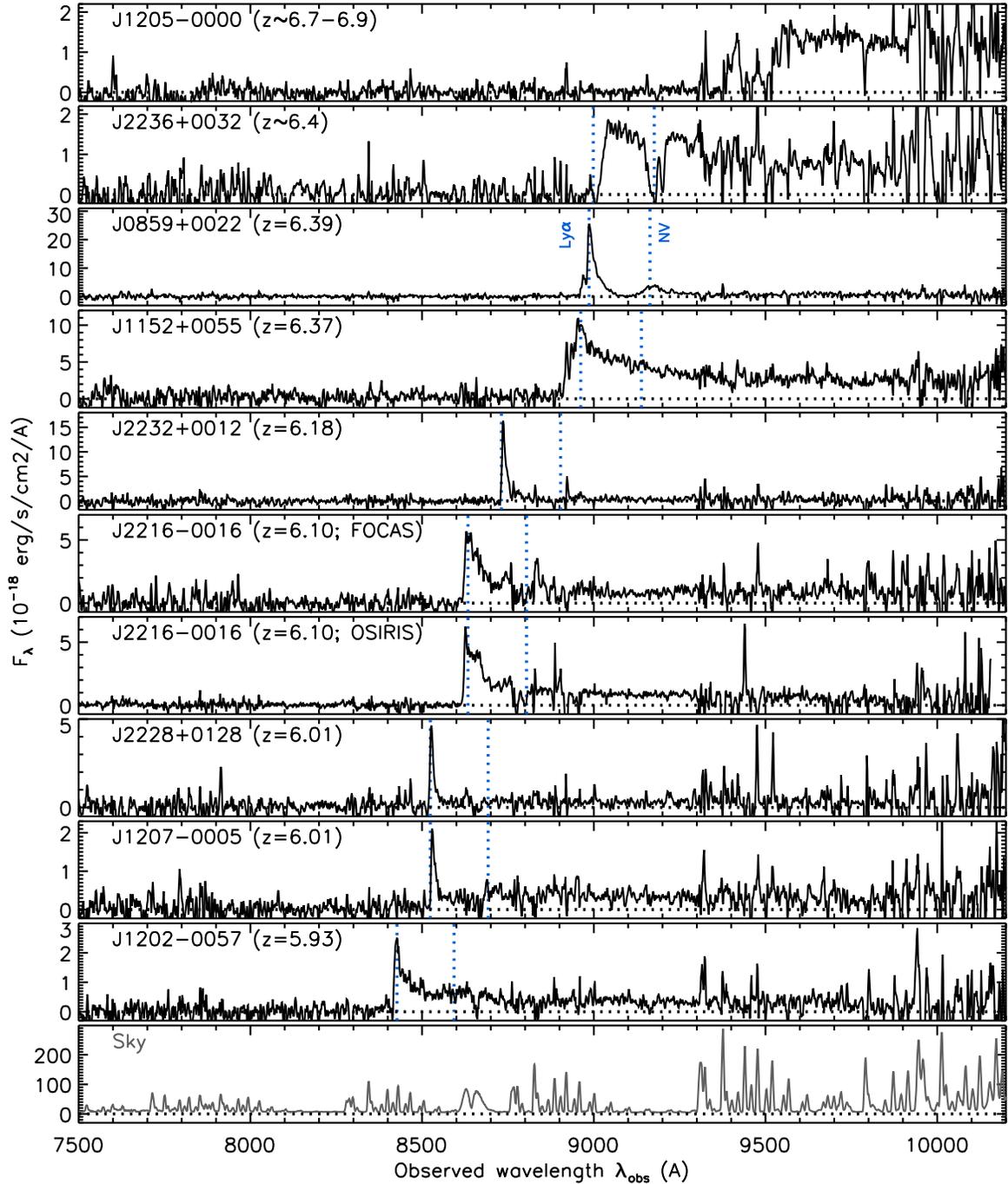}
\caption{The reduced spectra of the quasars and possible quasars discovered in this work, displayed in decreasing order of redshift.
The object name and the estimated redshift are indicated in the top-left corner of each panel.
The blue dotted lines mark the expected positions of the Ly$\alpha$ and \ion{N}{5} $\lambda$1240 emission lines, given the redshifts.
The spectra were smoothed using inverse-variance weighted means in 3 -- 7 pixel (depending on the S/N) boxes, for display purposes.
The bottom panel displays a sky spectrum.
\label{fig:spectra1}}
\end{figure*}

\begin{figure*}
\epsscale{1.0}
\plotone{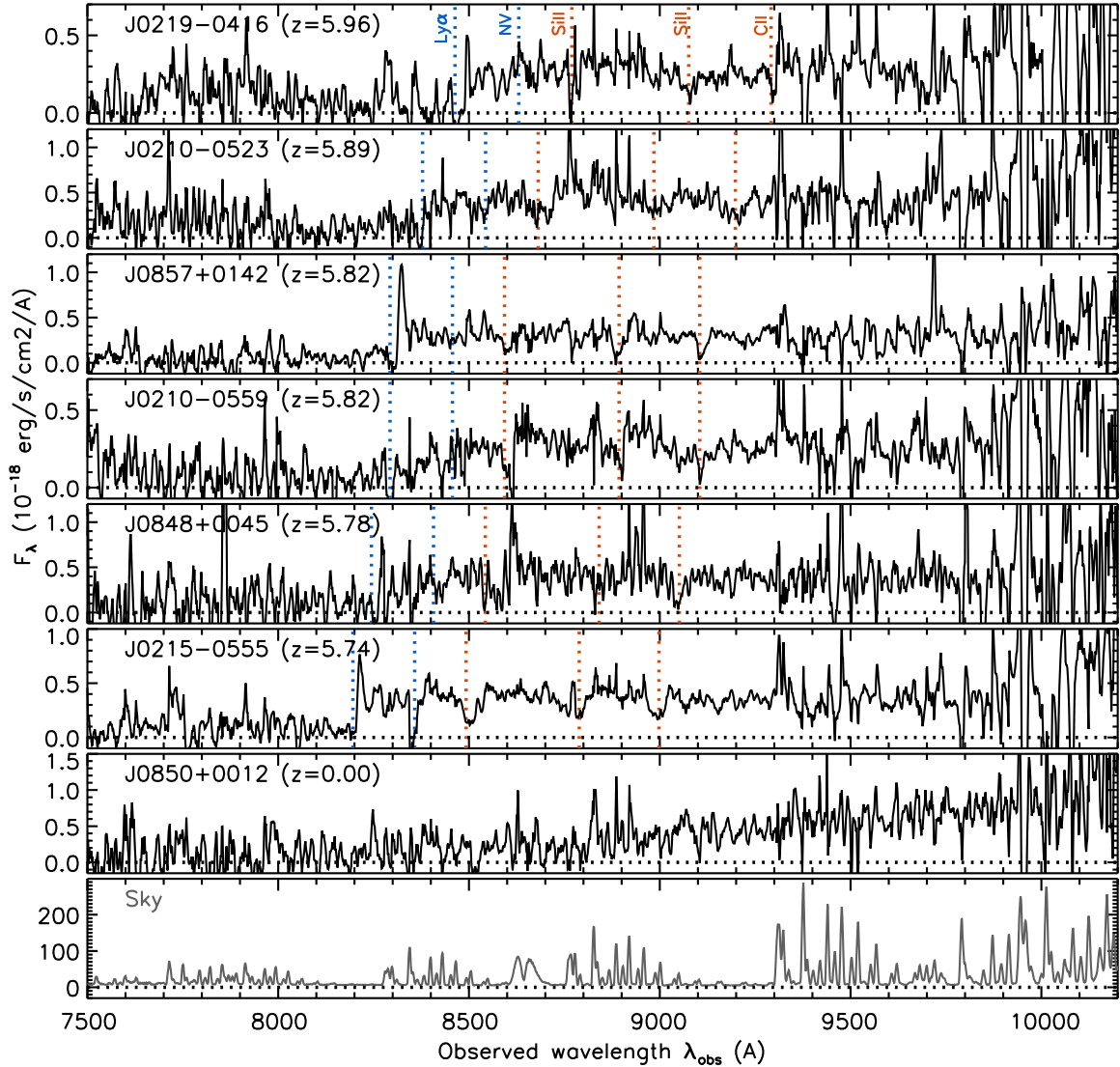}
\caption{Same as Figure \ref{fig:spectra1}, but for the high-$z$ galaxies (top six panels) and the brown dwarf (J0850$+$0012).
The expected positions of the interstellar absorption lines of \ion{Si}{2} $\lambda$1260, \ion{Si}{2} $\lambda$1304, and \ion{C}{2} $\lambda$1335 
are marked by the red dotted lines
for the galaxies.
\label{fig:spectra2}}
\end{figure*}

Figure \ref{fig:spectra1} displays the spectra of the identified quasars and possible quasars, while Figure \ref{fig:spectra2} displays 
those of the non-quasars (i.e., galaxies and a brown dwarf).
Their photometric and spectroscopic properties are summarized in Table \ref{tab:properties}.
We present short notes on the individual quasars in \S \ref{subsec:notes} and on the contaminating objects in  \S \ref{subsec:contaminants}. 
Discussion and future prospects are described in \S \ref{subsec:discussion}.

\begin{deluxetable*}{lrrccccccc}
\tabletypesize{\scriptsize}
\tablecaption{Spectroscopically Identified Objects \label{tab:properties}}
\tablewidth{0pt}
\tablehead{
\colhead{Name}  & \colhead{R.A.} & \colhead{Decl.} & \colhead{$i_{\rm AB}$ (mag)}  & \colhead{$z_{\rm AB}$ (mag)} & \colhead{$y_{\rm AB}$ (mag)} 
& \colhead{$J_{\rm AB}$ (mag)} & \colhead{$P_{\rm Q}^{\rm B}$} & \colhead{Redshift} 
& \colhead{$M_{1450}$ (mag)}
}
\startdata
\cutinhead{Quasars and Possible Quasars}
HSC J1205$-$0000  & 12:05:05.09  & $-$00:00:27.9   & $> 26.69$               & $> 26.45$               & 22.61 $\pm$ 0.03 & 21.95 $\pm$ 0.21 & 1.000 & 6.7--6.9 & $-$24.35 $\pm$ 0.18\\
HSC J2236$+$0032 & 22:36:44.58  & $+$00:32:56.9  & $> 27.05$               & 23.78 $\pm$ 0.08 & 23.23 $\pm$ 0.05 & \nodata                    &1.000 & 6.4    & $-$23.66 $\pm$ 0.10\\
HSC J0859$+$0022 & 08:59:07.19  & $+$00:22:55.9  & 27.89 $\pm$ 1.06 & 22.77 $\pm$ 0.01  & 23.65 $\pm$ 0.07 & $> 21.77$               & 1.000 & 6.39   & $-$23.59 $\pm$ 0.15\\
HSC J1152$+$0055 & 11:52:21.27  & $+$00:55:36.6  & 25.52 $\pm$ 0.09 & 21.83 $\pm$ 0.01  & 21.61 $\pm$ 0.02 & 21.66 $\pm$ 0.22  & 1.000 & 6.37   & $-$24.97 $\pm$ 0.11\\
HSC J2232$+$0012 & 22:32:12.03  &  $+$00:12:38.4 & 27.76 $\pm$ 0.51 & 23.84 $\pm$ 0.04  & 24.26 $\pm$ 0.13 & \nodata                    & 1.000 & 6.18   & $-$22.56 $\pm$ 0.34\\
HSC J2216$-$0016 & 22:16:44.47   & $-$00:16:50.1   & 26.05 $\pm$ 0.14 & 22.78 $\pm$  0.02 & 22.94 $\pm$ 0.03 & \nodata                    & 1.000 & 6.10   & $-$23.58 $\pm$ 0.12\\
HSC J2228$+$0128 & 22:28:27.83  & $+$01:28:09.5  & 27.56 $\pm$ 0.41 &24.06 $\pm$ 0.05   & 24.50 $\pm$ 0.13 & \nodata                    & 1.000 & 6.01   & $-$22.40 $\pm$ 0.12\\
HSC J1207$-$0005  & 12:07:54.14  & $-$00:05:53.3   & 26.39 $\pm$ 0.15 & 24.00 $\pm$ 0.03  & 23.87 $\pm$ 0.08 & $> 21.77$               & 1.000 & 6.01   &  $-$22.59 $\pm$ 0.08\\
HSC J1202$-$0057  & 12:02:46.37  & $-$00:57:01.7   & 26.13 $\pm$ 0.13 & 23.82 $\pm$ 0.03  & 23.89 $\pm$ 0.10 & $> 21.77$               & 1.000 & 5.93   & $-$22.54 $\pm$ 0.18\\
  \cutinhead{Galaxies}
HSC J0219$-$0416  & 02:19:29.41  & $-$04:16:45.9   &$> 26.74$                & 24.32 $\pm$ 0.06  & 24.14 $\pm$ 0.11& $> 21.78$                & 1.000 & 5.96  & $-$22.50 $\pm$ 0.07\\
HSC J0210$-$0523  & 02:10:33.82  &  $-$05:23:04.1  & 25.85 $\pm$ 0.17 & 23.78 $\pm$ 0.06  & 23.54 $\pm$ 0.11 & \nodata                    & 0.940 & 5.89  & $-$22.92 $\pm$ 0.20\\
HSC J0857$+$0142 & 08:57:23.95  & $+$01:42:54.6  & 26.24 $\pm$ 0.26 &  24.14 $\pm$ 0.05 & 23.93 $\pm$ 0.09 & $> 21.76$               & 0.964 & 5.82  & $-$22.52$\pm$ 0.03\\
HSC J0210$-$0559 & 02:10:41.28  &  $-$05:59:17.9  & 26.61 $\pm$ 0.27 & 24.25 $\pm$ 0.07  & 24.10 $\pm$ 0.16 & \nodata                    & 0.996 & 5.82  & $-$22.37 $\pm$ 0.06\\
HSC J0848$+$0045 & 08:48:18.33  &  $+$00:45:09.6 & 26.27 $\pm$ 0.21 & 23.90 $\pm$ 0.06  & 24.00 $\pm$ 0.10 & $>21.76$                & 1.000 & 5.78  & $-$22.75 $\pm$ 0.08\\
HSC J0215$-$0555  & 02:15:45.20  & $-$05:55:29.1   & 26.03 $\pm$ 0.15 & 23.98 $\pm$ 0.05  & 23.67 $\pm$ 0.10 & \nodata                    & 0.929 & 5.74  & $-$22.66 $\pm$ 0.02\\
\cutinhead{Brown dwarf}
HSC J0850$+$0012 & 08:50:02.63 & $+$00:12:10.0 & 28.40 $\pm$ 1.44 & 24.06 $\pm$ 0.06 & 23.22 $\pm$ 0.05 & $> 21.76$ & 0.171 & 0.00 & \nodata
\enddata
\tablecomments{The coordinates are at J2000.0. The $J$-band magnitudes are taken from VIKING. The magnitude upper limits are placed at 5$\sigma$ significance.
The errors of $M_{1450}$ do not include the uncertainty inherent in the assumed quasar and galaxy spectral slopes.}
\end{deluxetable*}

\subsection{Notes on individual quasars \label{subsec:notes}}

\subsubsection{HSC J1205$-$0000}

This object is a $z$-band dropout source with the HSC photometry.
While the deep IGM absorption trough at $\lambda_{\rm obs} < 9350$ \AA\ suggests that the redshift of this source is $z \ga 6.7$, the interpretation of the redder part 
of the spectrum is difficult with the present S/N.
Assuming that the Ly$\alpha$ line is somewhere in the spike around 9400 \AA, the inferred redshift is $z = 6.71 - 6.76$.
Then \ion{N}{5} $\lambda$1240 is expected at $\lambda_{\rm obs} = 9560 - 9620$ \AA, and the blueward absorption feature is likely to be an \ion{N}{5}
broad absorption line (BAL) system.
An alternative, interesting possibility 
is that  the strongest peak at $\lambda_{\rm obs} = 9550$ \AA\ is Ly$\alpha$ at $z = 6.85$.
In this case, the residual flux around 9400 \AA\  would imply that there is an ionized bubble within the IGM at $z \sim$ 6.7.
If this bubble is created by the ionizing radiation from J1205$-$0000, then the estimated near-zone size is $\sim$6 Mpc, or $\sim$13 Mpc when normalized
to $M_{1450} = -27$ mag, which is considerably larger than implied from other high-$z$ quasars \citep{carilli10}.
This problem would be alleviated if the bubble is ionized by a different source in front of J1205$-$0000, although the chance of having
such a high-$z$ foreground source must be small.
While the case for the redshift to be $z = 6.71 - 6.76$ is most likely, we need further data (e.g., other emission lines in deep near-IR or submm spectra) 
to determine an accurate redshift of this object.

\subsubsection{HSC J2236$+$0032} 

The IGM absorption trough at $\lambda_{\rm obs} < 9000$ \AA\ indicates that the redshift of this source is $z \ga 6.4$.
The redder part of the spectrum is relatively blue and indicates that this object is a quasar with no Ly$\alpha$ emission line.
The absence of Ly$\alpha$ may be due to the intrinsic nature of this quasar or by the damping wing of severe IGM absorption.
If we assume the Ly$\alpha$ wavelength of $\lambda_{\rm obs} = 9000$ \AA, then the estimated redshift is $z = 6.4$. 
In this case, the spectral position of the strong BAL observed at around 9200 \AA\ corresponds to \ion{N}{5} $\lambda$1240.
This BAL gas may also be responsible for absorbing the Ly$\alpha$ line emission.

\subsubsection{HSC J0859$+$0022}

This is an unambiguous quasar with strong and broad emission lines at $\lambda_{\rm obs} = 8990$ and 9170 \AA, which are
Ly$\alpha$ and \ion{N}{5} $\lambda$1240 at $z = 6.39$.
The continuum is very weak even at the wavelengths that are unaffected by IGM absorption (i.e., redward of the Ly$\alpha$ line),
and the observed broad-band flux in this spectral range is mostly contained in the above two lines.

\subsubsection{HSC J1152$+$0055}

The spectrum is typical of a high-$z$ quasar.
The redshift measured with the Ly$\alpha$ line is $z = 6.37$, but is quite uncertain, given the asymmetry in the line due to the onset of the Ly$\alpha$ forest.

\subsubsection{HSC J2232$+$0012}

The spectrum is reminiscent of those of Ly$\alpha$ emitters found in deep galaxy surveys \citep[e.g.,][]{ono12}.
The Ly$\alpha$ redshift is $z = 6.18$.
The luminosity, rest-frame equivalent width, and full width at half maximum (FWHM; after correcting for the instrumental broadening) 
of the line are $L ({\rm Ly}\alpha) \sim 10^{44.1}$ erg s$^{-1}$, $W_0 ({\rm Ly}\alpha) \sim 120$ \AA, and $v_{\rm FWHM} ({\rm Ly}\alpha) \sim 300$ km s$^{-1}$, respectively.
The high $L$ (Ly$\alpha$) implies a contribution from AGN,
as a recent study of a large sample of Ly$\alpha$ emitters at $z = 2.2$ \citep{konno16} suggests that the bright emitters with $L ({\rm Ly}\alpha) > 10^{43.4}$ erg s$^{-1}$ 
almost always have a substantial AGN contribution, based on their X-ray, UV, and radio properties.
The measured $L$ (Ly$\alpha$) and $v_{\rm FWHM} ({\rm Ly}\alpha)$ are similar to those of the high-$z$ galaxy ``CR7", which is suggested to contain 
Population-III-like stars \citep[][based on the very strong nebular lines including \ion{He}{2} $\lambda$1640]{sobral15}, while $W_0 ({\rm Ly}\alpha)$ 
is less than half that of CR7.
The AGN contribution in this object may also be supported by the possible broad-line component seen redward of Ly$\alpha$, 
as well as the relatively large $v_{\rm FWHM} ({\rm Ly}\alpha)$ \citep[see, e.g.,][]{alexandroff13}.
Note that the intrinsic line FWHM may be as much as twice the value estimated here, due to the IGM absorption.

\subsubsection{HSC J2216$-$0016}

This is an unambiguous quasar at $z = 6.10$, observed both at GTC and Subaru. The spectrum around the expected \ion{N}{5} $\lambda$1240 position
($\lambda_{\rm obs} = 8790$ \AA) is heavily absorbed, likely due to a BAL of the \ion{N}{5} line.
The small-scale features of this BAL are very similar in the FOCAS and OSIRIS spectra, suggesting that they are real.
We note that there is no atmospheric absorption feature in the standard star spectrum at the wavelengths corresponding to this BAL.

\subsubsection{HSC J2228$+$0128}

The redshift measured with Ly$\alpha$ is $z = 6.01$.
The spectrum is similar to but not as extreme as J2232$+$0012, with $L ({\rm Ly}\alpha) \sim 10^{43.3}$ erg s$^{-1}$, $W_0 ({\rm Ly}\alpha) \sim 20$ \AA, and 
$v_{\rm FWHM} ({\rm Ly}\alpha) \sim 270$ km s$^{-1}$.
As in J2232$+$0012, the luminous Ly$\alpha$ line implies a contribution from AGN. 

\subsubsection{HSC J1207$-$0005} 

The spectrum is very similar to that of the previous object, J2228$+$0128.
The redshift inferred from the Ly$\alpha$ line is $z = 6.01$, while the Ly$\alpha$ properties are 
$L ({\rm Ly}\alpha) \sim 10^{43.0}$ erg s$^{-1}$, $W_0 ({\rm Ly}\alpha) \sim 10$ \AA, and $v_{\rm FWHM} ({\rm Ly}\alpha) \sim 420$ km s$^{-1}$.
The luminous Ly$\alpha$ emission and relatively large $v_{\rm FWHM} ({\rm Ly}\alpha)$ imply an AGN contribution, as in J2232$+$0012 and J2228$+$0128.

\subsubsection{HSC J1202$-$0057}

This object has a typical quasar spectrum, with a strong and broad Ly$\alpha$ line and blue continuum. 
The redshift estimated with Ly$\alpha$ is $z = 5.93$.

\subsection{Notes on the contaminating objects \label{subsec:contaminants}}

Figure \ref{fig:spectra2} presents the spectra of the objects that are not likely to be quasars, based on the absence of high ionization lines,
broad emission lines, or blue continuum.
J0857$+$0142 and J0215$-$0555 have narrow Ly$\alpha$ lines and sharp continuum breaks characteristic of GP troughs.
The interstellar absorption lines of  \ion{Si}{2} $\lambda$1260, \ion{Si}{2} $\lambda$1304, and \ion{C}{2} $\lambda$1335 are also clearly visible,
which indicate that they are LBGs at $z = 5.89$ and $z = 5.74$, respectively.
These three absorption lines may also be present in the spectra of J0219$-$0416, J0210$-$0523, J0210$-$0559, and J0848$+$0045, although the S/N is lower.
Combined with the relatively flat continua and the sharp flux drops in the bluer part of the spectra, they are also likely to be high-$z$ galaxies.
Their approximate redshifts were estimated from the absorption lines.

It is worth mentioning that the redshift distribution of these galaxies ($5.7 < z < 6.0$) is systematically different from that of the discovered quasars ($5.9 < z < 6.9$).
This is partly because only intrinsically luminous objects can be detected at $z > 6.0$, where the GP trough comes into the $z$-band.
In addition, galaxies are redder than quasars at $6.0 < z < 6.3$ due to weaker Ly$\alpha$ lines and redder continua, and so are more difficult
to separate from the Galactic brown dwarfs (see Figure \ref{fig:colordiagram}).
The evolution of the luminous end of the galaxy luminosity function around $z \sim 6$ may also be a factor.

Another interpretation is that these sources are red galaxies at $z \sim 1$, whose 4000 \AA\ break is responsible for their red $i - z$ colors 
(see the discussion in \S 3).
However, we find this to be unlikely.
Even with high metallicity (2.5 $Z_\odot$) and old stellar age (5 Gyr), a large amount of dust reddening ($E_{\rm B-V} > 1.5$) is required
for a galaxy at $z \sim 1$ to have $i_{\rm AB} - z_{\rm AB} > 1.5$ \citep{toshikawa12}.
Such high values of $E_{\rm B-V}$ and $i_{\rm AB} - z_{\rm AB}$ are rarely found in samples of $z \sim 1$ red galaxies \citep[e.g.,][]{miyazaki03, malhotra05},
while our objects have even redder colors ($i_{\rm AB} - z_{\rm AB} > 2.0$). 
In addition, their spectra show absorption lines in most of the cases consistent with \ion{Si}{2} $\lambda$1260, \ion{Si}{2} $\lambda$1304, 
and \ion{C}{2} $\lambda$1335 at $z > 5.7$.
The lack of lower-$z$ galaxies in our sample demonstrates the power of HSC, whose excellent image quality helps detect faint extended emission of such 
galaxies and separate them from high-$z$ objects morphologically.

Finally, J0850$+$0012 has a relatively smooth red continuum characteristic of a brown dwarf.
Indeed this object has a rather low $P_{\rm Q}^{\rm B}$ value ($P_{\rm Q}^{\rm B} = 0.17$) and the Bayesian algorithm predicts that it is most likely a L0 dwarf,
which is consistent with the obtained spectrum.
A follow-up study of this faint dwarf will be presented in a future paper.

\subsection{Discussion and future prospects \label{subsec:discussion}}

In summary, we obtained spectra of 19 out of the 38 candidates, and identified 15 new high-$z$ quasars and galaxies.
The above candidates also include 5 previously known quasars, which were recovered by our selection.
The current status of the spectroscopic identifications is presented in Figure \ref{fig:maghistogram}.
Our selection is quite successful at magnitudes brighter than $z_{\rm AB} = 23.5$ mag, where seven out of the nine candidates have turned out to be quasars and 
the remaining two are awaiting spectroscopy.
This implies the following two points:
First, the source detection and measurements with the HSC hardware and reduction software are highly reliable without serious systematic effects.
Second, our quasar selection algorithms work quite efficiently as long as they are fed with correct photometry information.

\begin{figure}
\epsscale{1.1}
\plotone{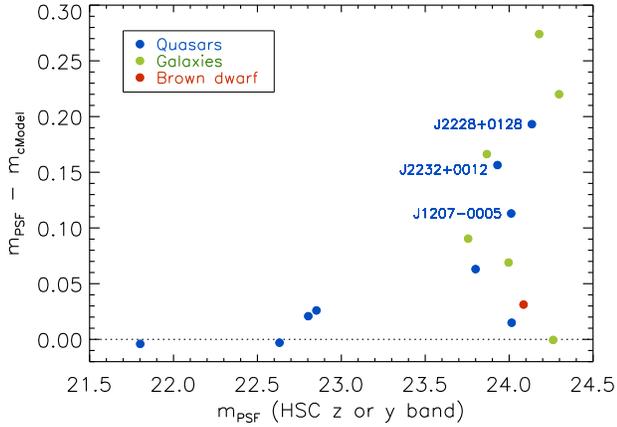}
\caption{The difference between the PSF magnitude ($m_{\rm PSF}$) and cModel magnitude ($m_{\rm cModel}$) as a function of $m_{\rm PSF}$, 
for quasars and possible quasars (blue), galaxies (green), and a brown dwarf (red).
The magnitudes are measured in the $y$-band for the $z \sim 6.8$ quasar, J1205$-$0000, and in the $z$-band for the remaining objects.
Because cModel magnitudes in our fiducial (S15A) data release are known to be biased, we use magnitudes from the newer (S15B) release
only in this plot.
The three quasars with relatively large $m_{\rm PSF} - m_{\rm cModel}$ values are  labeled with their names.
\label{fig:extendedness}}
\end{figure}

Interestingly, we have started to find high-$z$ galaxies as significant contaminants at $z_{\rm AB} > 23.5$ mag.
This is not surprising because the luminosity functions of quasars and LBGs are likely to intersect at $\sim$24 mag, with LBGs outnumbering quasars at fainter magnitudes 
(see Figure \ref{fig:numbercounts} and the discussion in \S 3).
Figure \ref{fig:extendedness} presents the difference between the PSF and cModel magnitudes, as a measure of source extendedness,
for all the identified objects in this work. 
Because the host galaxy contribution is not always negligible for the low-luminosity quasars we are looking for, 
and the extendedness is a noisy quantity at the faintest magnitudes,
no clear cut 
can be defined to separate quasars and galaxies on this plane.
Indeed, the three possible quasars with relatively narrow Ly$\alpha$ lines (J2232$+$0012, J2228$+$0128, J1207$-$0005; see Figure \ref{fig:spectra1})
appear to have larger extendedness than the remaining quasars, suggesting significant light from the host galaxies.

Meanwhile, the discovered galaxies are an important probe of the reionization era. 
For example, stellar populations in such high-$z$ bright galaxies can be studied in detail with high-quality spectra. 
Measurements of the interstellar absorption lines such as those observed in J0857$+$0142 and J0215$-$0555 have the potential to constrain 
the escape fraction of ionizing photons \citep{jones13}.
Follow-up observations with facilities at other wavelengths (e.g., the Atacama Large Millimeter/submillimeter Array; ALMA), would also be useful to 
understand the nature of these galaxies in the high-$z$ Universe.

At the faintest magnitudes of our survey ($24.0 < z_{\rm AB} < 24.5$ mag), the photometric selection and spectroscopic identification become more challenging.
Along with the increasing fraction of galaxies, we found a contaminating brown dwarf, which is expected from its low Bayesian quasar probability 
($P_{\rm Q}^{\rm B}$ = 0.17; see \S \ref{subsec:contaminants}).
We were not able to confirm the nature of two candidates due to their low spectral S/N, although we typically spent a few hours per object with Subaru.
They seem to have no strong emission lines and so may be galaxies or brown dwarfs, but weak-line quasars such as J2236$+$0032 are known to exist.
Further analysis of these objects, possibly with additional observing time to increase the S/N, will be presented in a future paper.

With the five previously known objects and the nine newly identified objects (including the possible quasars J2232$+$0012, J2228$+$0128, and J1207$-$0005), 
fourteen high-$z$ quasars 
are now known in the present survey area. 
This is roughly 
half of 
the expected number with our survey limit in 80 deg$^2$ (see \S 2),
although the spectroscopic identification is still not complete.
In the last column of Table \ref{tab:properties}, we report the absolute magnitudes of the discovered quasars as well as galaxies.
For quasars, we measure the flux densities at the rest-frame wavelength $\lambda_{\rm rest} = 1270 - 1330$ \AA\ (except for J$1205-0000$ with the
highest redshift, in which case we adopt $\lambda_{\rm rest} = 1250 - 1260$ \AA\ and assume $z = 6.75$) and estimate $M_{\rm 1450}$ assuming the power-law 
continuum slope $\alpha = -1.5$ 
\citep[$f_\lambda \propto \lambda^\alpha$; e.g.,][]{vandenberk01}.
For galaxies, we measure $M_{1350}$ at $\lambda_{\rm rest} = 1320 - 1380$ \AA\ and convert them to $M_{1450}$ by assuming the UV spectral slope of $\beta = -2$
\citep{stanway05}.
As expected, our survey has succeeded in identifying quasars (and galaxies) with luminosities approaching $M_{1450} \sim -22$ mag (i.e., $\sim$2 mag lower 
luminosity than found in most of the previous large surveys; see \S 1).


We will continue to develop the SHELLQs project as the HSC-SSP survey continues.
The present work only partially covers the first 80 deg$^2$ of the Wide layer, which will eventually observe 1400 deg$^2$.
Our immediate goal is to complete the spectroscopic identification in this 80 deg$^2$ area and derive our first quasar luminosity function.
In the long term, we will expand the survey area and significantly increase the sample size and luminosity range of known high-$z$ quasars. 
As described in \S 1, the expected numbers of quasars over the whole Wide area are $\sim$500 with $z_{\rm AB} < 24.5$ mag at $z \sim 6$ and 
$\sim$100 with $y_{\rm AB} < 24.0$ mag at $z \sim 7$.
We will also explore even lower luminosities with the Deep and Ultra-Deep layers of the HSC-SSP survey, although follow-up spectroscopy will become more challenging.

At the same time, it is important to follow up individual quasars in greater detail.
The redshifts of some of the discovered quasars are poorly constrained at the moment, which should be improved.
We are also planning deep optical and NIR spectroscopy to measure the near-zone size, SMBH mass, and metallicity for those quasars at lower luminosity
than previously known at $z > 6$.
These low-luminosity objects are expected to be much more numerous than the brighter ones, and hence possess critical information about the general
properties of quasars in the early Universe.
They will also provide a useful constraint on the low-mass end of the SMBH mass function, and in turn, models of the formation and early evolution of SMBHs.
In addition, X-ray observations will play a critical role to estimate the bolometric luminosity, the Eddington ratio, and the presence and properties
of absorbing material.
We also plan to conduct ALMA follow-up observations in order to study the gas and dust content, as well as the star formation activity in the host galaxies.

\section{Summary \label{sec:summary}}

We present initial results from the SHELLQs project, a survey of low-luminosity quasars and AGNs at high redshift close to the reionization era.
The project exploits the exquisite imaging data with five optical bands  ($g$, $r$, $i$, $z$, and $y$) produced by the Subaru HSC-SSP survey, supplemented
with NIR photometry where available from UKIDSS and VIKING.
The limiting magnitudes of the quasar search are currently set to $z_{\rm AB} < 24.5$ mag and $y_{\rm AB} < 24.0$ mag, but these may change in the future.
The candidates are selected by combining several photometric approaches, including a Bayesian probabilistic algorithm, which have turned out to be
quite efficient in eliminating astrophysical contaminants such as stars and dwarfs, as well as cosmic rays, moving objects, and transient events.
From the early HSC-SSP survey area covering 80 deg$^2$, we identified 38 candidate high-$z$ quasars, which are the focus of this paper.

We carried out spectroscopic follow-up observations of 19 of these candidates, with GTC/OSIRIS and Subaru/FOCAS, in the 2015 Fall and 2016 Spring semesters.
Nine objects were identified as quasars or possible quasars at $5.9 < z < 6.9$, based on the sharp continuum breaks characteristic of GP troughs,
broad Ly$\alpha$ and \ion{N}{5} $\lambda$1240 lines, and/or blue continuum.
Six objects are likely high-$z$ galaxies with interstellar absorption lines of \ion{Si}{2} $\lambda$1260, \ion{Si}{2} $\lambda$1304, and \ion{C}{2} $\lambda$1335,
and in some cases narrow Ly$\alpha$ emission lines.
The remaining objects include a L0 dwarf, a moving or transient object, and two sources whose nature is still uncertain due to the low spectral S/N.
In addition to these newly identified objects, five quasars were known prior to our survey among the 38 candidates.
The success rate of our selection is quite high, and most of the objects we took spectra of were identified as high-$z$ quasars or galaxies.

The SHELLQs project will continue as the HSC-SSP survey continues toward its goals of observing 1400 deg$^2$ in the Wide layer,
as well as 27 and 3.5 deg$^2$ in the Deep and Ultra-Deep layer, respectively.
We will soon deliver our first quasar luminosity function reaching down to $M_{\rm AB} \sim -22$ mag at $z \sim 6$.
Further follow-up observations of the discovered quasars and galaxies are being considered at various wavelengths from sub-millimeter/radio to X-ray.

\acknowledgments

We thank the referee for his/her useful comments and suggestions to improve this paper.
We are grateful to everyone involved in the hardware development, observations, and data reduction for the HSC-SSP survey.
We had a lot of great help from Chien-Hsiu Lee, Takashi Hattori, and other Subaru staff members for the FOCAS observations.
NK acknowledges support from the Japan Society for the Promotion of Science (JSPS) through Grant-in-Aid for Scientific Research 15H03645.
KI acknowledges support by the Spanish MINECO under grant AYA2013-47447-C3-2-P and MDM-2014-0369 of ICCUB (Unidad de Excelencia `Mar\'{i}a de Maeztu').
TN acknowledges financial support from the JSPS (KAKENHI grant no. 25707010) and also from the JGC-S Scholarship Foundation.

The HSC collaboration includes the astronomical
communities of Japan and Taiwan, and Princeton University.  The HSC
instrumentation and software were developed by the National
Astronomical Observatory of Japan (NAOJ), the Kavli Institute for the
Physics and Mathematics of the Universe (Kavli IPMU), the University
of Tokyo, the High Energy Accelerator Research Organization (KEK), the
Academia Sinica Institute for Astronomy and Astrophysics in Taiwan
(ASIAA), and Princeton University.  Funding was contributed by the FIRST 
program from Japanese Cabinet Office, the Ministry of Education, Culture, 
Sports, Science and Technology (MEXT), the Japan Society for the 
Promotion of Science (JSPS),  Japan Science and Technology Agency 
(JST),  the Toray Science  Foundation, NAOJ, Kavli IPMU, KEK, ASIAA,  
and Princeton University.

This paper makes use of software developed for the LSST. We thank the LSST Project for 
making their code available as free software at http://dm.lsstcorp.org.

The Pan-STARRS1 Surveys (PS1) have been made possible through contributions of the Institute for Astronomy, the University of Hawaii, the Pan-STARRS Project Office, the Max-Planck Society and its participating institutes, the Max Planck Institute for Astronomy, Heidelberg and the Max Planck Institute for Extraterrestrial Physics, Garching, The Johns Hopkins University, Durham University, the University of Edinburgh, Queen's University Belfast, the Harvard-Smithsonian Center for Astrophysics, the Las Cumbres Observatory Global Telescope Network Incorporated, the National Central University of Taiwan, the Space Telescope Science Institute, the National Aeronautics and Space Administration under Grant No. NNX08AR22G issued through the Planetary Science Division of the NASA Science Mission Directorate, the National Science Foundation under Grant No. AST-1238877, the University of Maryland, Eotvos Lorand University (ELTE), and the Los Alamos National Laboratory.



{\it Facilities:} \facility{Subaru}, \facility{GTC}.





\begin{thebibliography}{}
\bibitem[Abazajian et al.(2004)]{abazajian04} Abazajian, K., Adelman-McCarthy, J.~K., Ag{\"u}eros, M.~A., et al.\ 2004, \aj, 128, 502 
\bibitem[Alexandroff et al.(2013)]{alexandroff13} Alexandroff, R., Strauss, M.~A., Greene, J.~E., et al.\ 2013, \mnras, 435, 3306 
\bibitem[Axelrod et al.(2010)]{axelrod10} Axelrod, T., Kantor, J., Lupton, R.~H., \& Pierfederici, F.\ 2010, \procspie, 7740, 774015 
\bibitem[Ba{\~n}ados et al.(2014)]{banados14} Ba{\~n}ados, E., Venemans, B.~P., Morganson, E., et al.\ 2014, \aj, 148, 14 
\bibitem[Bertin \& Arnouts(1996)]{bertin96} Bertin, E., \& Arnouts, S.\ 1996, \aaps, 117, 393 
\bibitem[Bouwens et al.(2015)]{bouwens15} Bouwens, R.~J., Illingworth, G.~D., Oesch, P.~A., et al.\ 2015, \apj, 803, 34 
\bibitem[Bowler et al.(2015)]{bowler15} Bowler, R.~A.~A., Dunlop, J.~S., McLure, R.~J., et al.\ 2015, \mnras, 452, 1817 
\bibitem[Bowler et al.(2014)]{bowler14} Bowler, R.~A.~A., Dunlop, J.~S., McLure, R.~J., et al.\ 2014, \mnras, 440, 2810 
\bibitem[Carnall et al.(2015)]{carnall15} Carnall, A.~C., Shanks, T., Chehade, B., et al.\ 2015, \mnras, 451, L16 
\bibitem[Caballero et al.(2008)]{caballero08} Caballero, J.~A., Burgasser, A.~J., \& Klement, R.\ 2008, \aap, 488, 181 
\bibitem[Carilli et al.(2010)]{carilli10} Carilli, C.~L., Wang, R., Fan, X., et al.\ 2010, \apj, 714, 834 
\bibitem[Cepa et al.(2000)]{cepa00} Cepa, J., Aguiar, M., Escalera, V.~G., et al.\ 2000, \procspie, 4008, 623 
\bibitem[Choudhury et al.(2015)]{choudhury15} Choudhury, T.~R., Puchwein, E., Haehnelt, M.~G., \& Bolton, J.~S.\ 2015, \mnras, 452, 261 
\bibitem[Dark Energy Survey Collaboration(2016)]{des16} Dark Energy Survey Collaboration, Abbott, T., Abdalla, F.~B., et al.\ 2016, arXiv:1601.00329 
\bibitem[De Rosa et al.(2011)]{derosa11} De Rosa, G., Decarli, R., Walter, F., et al.\ 2011, \apj, 739, 56 
\bibitem[De Rosa et al.(2014)]{derosa14} De Rosa, G., Venemans, B.~P., Decarli, R., et al.\ 2014, \apj, 790, 145 
\bibitem[Fan et al.(2006a)]{fan06araa} Fan, X., Carilli, C.~L., \& Keating, B.\ 2006, \araa, 44, 415 
\bibitem[Fan et al.(2004)]{fan04} Fan, X., Hennawi, J.~F., Richards, G.~T., et al.\ 2004, \aj, 128, 515 
\bibitem[Fan et al.(2001a)]{fan01} Fan, X., Narayanan, V.~K., Lupton, R.~H., et al.\ 2001, \aj, 122, 2833 
\bibitem[Fan et al.(2006b)]{fan06} Fan, X., Strauss, M.~A., Richards, G.~T., et al.\ 2006, \aj, 131, 1203 
\bibitem[Fan et al.(2003)]{fan03} Fan, X., Strauss, M.~A., Schneider, D.~P., et al.\ 2003, \aj, 125, 1649 
\bibitem[Fan et al.(2000)]{fan00} Fan, X., White, R.~L., Davis, M., et al.\ 2000, \aj, 120, 1167 
\bibitem[Ferrara et al.(2014)]{ferrara14} Ferrara, A., Salvadori, S., Yue, B., \& Schleicher, D.\ 2014, \mnras, 443, 2410 
\bibitem[Fontanot et al.(2012)]{fontanot12} Fontanot, F., Cristiani, S., \& Vanzella, E.\ 2012, \mnras, 425, 1413 
\bibitem[Fukugita et al.(1996)]{fukugita96} Fukugita, M., Ichikawa, T., Gunn, J.~E., et al.\ 1996, \aj, 111, 1748 
\bibitem[Furusawa et al.(2008)]{furusawa08} Furusawa, H., Kosugi, G., Akiyama, M., et al.\ 2008, \apjs, 176, 1 
\bibitem[Giallongo et al.(2015)]{giallongo15} Giallongo, E., Grazian, A., Fiore, F., et al.\ 2015, \aap, 578, A83 
\bibitem[Gonz{\'a}lez et al.(2012)]{gonzalez12} Gonz{\'a}lez, V., Bouwens, R.~J., Labb{\'e}, I., et al.\ 2012, \apj, 755, 148 
\bibitem[Goto(2006)]{goto06} Goto, T.\ 2006, \mnras, 371, 769 
\bibitem[Goto et al.(2009)]{goto09} Goto, T., Utsumi, Y., Furusawa, H., Miyazaki, S., \& Komiyama, Y.\ 2009, \mnras, 400, 843 
\bibitem[Goto et al.(2012)]{goto12} Goto, T., Utsumi, Y., Walsh, J.~R., et al.\ 2012, \mnras, 421, L77 
\bibitem[Grissom et al.(2014)]{grissom14} Grissom, R.~L., Ballantyne, D.~R., \& Wise, J.~H.\ 2014, \aap, 561, A90 
\bibitem[Gunn \& Peterson(1965)]{gunn65} Gunn, J.~E., \& Peterson, B.~A.\ 1965, \apj, 142, 1633 
\bibitem[Hewett et al.(2006)]{hewett06} Hewett, P.~C., Warren, S.~J., Leggett, S.~K., \& Hodgkin, S.~T.\ 2006, \mnras, 367, 454 
\bibitem[Ivezic et al.(2008)]{ivezic08} Ivezic, Z., Tyson, J.~A., Abel, B., et al.\ 2008, arXiv:0805.2366 
\bibitem[Jiang et al.(2008)]{jiang08} Jiang, L., Fan, X., Annis, J., et al.\ 2008, \aj, 135, 1057 
\bibitem[Jiang et al.(2009)]{jiang09} Jiang, L., Fan, X., Bian, F., et al.\ 2009, \aj, 138, 305 
\bibitem[Jiang et al.(2015)]{jiang15} Jiang, L., McGreer, I.~D., Fan, X., et al.\ 2015, \aj, 149, 188 
\bibitem[Jiang et al.(2007)]{jiang07} Jiang, L., Fan, X., Vestergaard, M., et al.\ 2007, \aj, 134, 1150 
\bibitem[Jones et al.(2013)]{jones13} Jones, T.~A., Ellis, R.~S., Schenker, M.~A., \& Stark, D.~P.\ 2013, \apj, 779, 52 
\bibitem[Juri{\'c} et al.(2015)]{juric15} Juri{\'c}, M., Kantor, J., Lim, K., et al.\ 2015, arXiv:1512.07914 
\bibitem[Kashikawa et al.(2002)]{kashikawa02} Kashikawa, N., Aoki, K., Asai, R., et al.\ 2002, \pasj, 54, 819 
\bibitem[Kashikawa et al.(2015)]{kashikawa15} Kashikawa, N., Ishizaki, Y., Willott, C.~J., et al.\ 2015, \apj, 798, 28 
\bibitem[Kashikawa et al.(2007)]{kashikawa07} Kashikawa, N., Kitayama, T., Doi, M., et al.\ 2007, \apj, 663, 765 
\bibitem[Kaiser et al.(2010)]{kaiser10} Kaiser, N., Burgett, W., Chambers, K., et al.\ 2010, \procspie, 7733, 77330E 
\bibitem[Kawaguchi et al.(2004)]{kawaguchi04} Kawaguchi, T., Aoki, K., Ohta, K., \& Collin, S.\ 2004, \aap, 420, L23 
\bibitem[Kim et al.(2015)]{kim15} Kim, Y., Im, M., Jeon, Y., et al.\ 2015, \apjl, 813, L35 
\bibitem[Konno et al.(2016)]{konno16} Konno, A., Ouchi, M., Nakajima, K., et al.\ 2015, arXiv:1512.01854 
\bibitem[Konno et al.(2014)]{konno14} Konno, A., Ouchi, M., Ono, Y., et al.\ 2014, \apj, 797, 16 
\bibitem[Lawrence et al.(2007)]{lawrence07} Lawrence, A., Warren, S.~J., Almaini, O., et al.\ 2007, \mnras, 379, 1599 
\bibitem[Lehnert \& Bremer(2003)]{lehnert03} Lehnert, M.~D., \& Bremer, M.\ 2003, \apj, 593, 630 
\bibitem[Madau \& Haardt(2015)]{madau15} Madau, P., \& Haardt, F.\ 2015, \apjl, 813, L8 
\bibitem[Madau et al.(2014)]{madau14} Madau, P., Haardt, F., \& Dotti, M.\ 2014, \apjl, 784, L38 
\bibitem[Magnier et al.(2013)]{magnier13} Magnier, E.~A., Schlafly, E., Finkbeiner, D., et al.\ 2013, \apjs, 205, 20 
\bibitem[Maiolino et al.(2005)]{maiolino05} Maiolino, R., Cox, P., Caselli, P., et al.\ 2005, \aap, 440, L51 
\bibitem[Malhotra et al.(2005)]{malhotra05} Malhotra, S., Rhoads, J.~E., Pirzkal, N., et al.\ 2005, \apj, 626, 666 
\bibitem[Miyazaki et al.(2012)]{miyazaki12} Miyazaki, S., Komiyama, Y., Nakaya, H., et al.\ 2012, \procspie, 8446, 84460Z 
\bibitem[Miyazaki et al.(2003)]{miyazaki03} Miyazaki, M., Shimasaku, K., Kodama, T., et al.\ 2003, \pasj, 55, 1079 
\bibitem[Mortlock et al.(2012)]{mortlock12} Mortlock, D.~J., Patel, M., Warren, S.~J., et al.\ 2012, \mnras, 419, 390 
\bibitem[Mortlock et al.(2011)]{mortlock11} Mortlock, D.~J., Warren, S.~J., Venemans, B.~P., et al.\ 2011, \nat, 474, 616 
\bibitem[Oke \& Gunn(1983)]{oke83} Oke, J.~B., \& Gunn, J.~E.\ 1983, \apj, 266, 713 
\bibitem[Ono et al.(2012)]{ono12} Ono, Y., Ouchi, M., Mobasher, B., et al.\ 2012, \apj, 744, 83 
\bibitem[Ouchi et al.(2008)]{ouchi08} Ouchi, M., Shimasaku, K., Akiyama, M., et al.\ 2008, \apjs, 176, 301 
\bibitem[Ouchi et al.(2010)]{ouchi10} Ouchi, M., Shimasaku, K., Furusawa, H., et al.\ 2010, \apj, 723, 869 
\bibitem[Planck Collaboration (2016)]{planck16} Planck Collaboration\ 2016, arXiv:1605.03507 
\bibitem[Pickles(1998)]{pickles98} Pickles, A.~J.\ 1998, \pasp, 110, 863 
\bibitem[Reed et al.(2015)]{reed15} Reed, S.~L., McMahon, R.~G., Banerji, M., et al.\ 2015, \mnras, 454, 3952 
\bibitem[Richards et al.(2002)]{richards02} Richards, G.~T., Fan, X., Newberg, H.~J., et al.\ 2002, \aj, 123, 2945 
\bibitem[Robertson et al.(2010)]{robertson10} Robertson, B.~E., Ellis, R.~S., Dunlop, J.~S., McLure, R.~J., \& Stark, D.~P.\ 2010, \nat, 468, 49 
\bibitem[Robertson et al.(2015)]{robertson15} Robertson, B.~E., Ellis, R.~S., Furlanetto, S.~R., \& Dunlop, J.~S.\ 2015, \apjl, 802, L19 
\bibitem[Robertson et al.(2013)]{robertson13} Robertson, B.~E., Furlanetto, S.~R., Schneider, E., et al.\ 2013, \apj, 768, 71 
\bibitem[Schlafly et al.(2012)]{schlafly12} Schlafly, E.~F., Finkbeiner, D.~P., Juri{\'c}, M., et al.\ 2012, \apj, 756, 158 
\bibitem[Schlegel et al.(1998)]{schlegel98} Schlegel, D.~J., Finkbeiner, D.~P., \& Davis, M.\ 1998, \apj, 500, 525 
\bibitem[Shapiro(2005)]{shapiro05} Shapiro, S.~L.\ 2005, \apj, 620, 59 
\bibitem[Sobral et al.(2015)]{sobral15} Sobral, D., Matthee, J., Darvish, B., et al.\ 2015, \apj, 808, 139 
\bibitem[Songaila(2004)]{songaila04} Songaila, A.\ 2004, \aj, 127, 2598 
\bibitem[Stanway et al.(2005)]{stanway05} Stanway, E.~R., McMahon, R.~G., \& Bunker, A.~J.\ 2005, \mnras, 359, 1184 
\bibitem[Tonry et al.(2012)]{tonry12} Tonry, J.~L., Stubbs, C.~W., Lykke, K.~R., et al.\ 2012, \apj, 750, 99 
\bibitem[Toshikawa et al.(2012)]{toshikawa12} Toshikawa, J., Kashikawa, N., Ota, K., et al.\ 2012, \apj, 750, 137 
\bibitem[Vanden Berk et al.(2001)]{vandenberk01} Vanden Berk, D.~E., Richards, G.~T., Bauer, A., et al.\ 2001, \aj, 122, 549 
\bibitem[Venemans et al.(2013)]{venemans13} Venemans, B.~P., Findlay, J.~R., Sutherland, W.~J., et al.\ 2013, \apj, 779, 24 
\bibitem[Venemans et al.(2012)]{venemans12} Venemans, B.~P., McMahon, R.~G., Walter, F., et al.\ 2012, \apjl, 751, L25 
\bibitem[Venemans et al.(2015b)]{venemans15b} Venemans, B.~P., Verdoes Kleijn, G.~A., Mwebaze, J., et al.\ 2015, \mnras, 453, 2259 
\bibitem[Venemans et al.(2016)]{venemans16} Venemans, B.~P., Walter, F., Zschaechner, L., et al.\ 2016, \apj, 816, 37 
\bibitem[Volonteri(2012)]{volonteri12} Volonteri, M.\ 2012, Science, 337, 544 
\bibitem[Wang et al.(2007)]{wang07} Wang, R., Carilli, C.~L., Beelen, A., et al.\ 2007, \aj, 134, 617 
\bibitem[Wang et al.(2013)]{wang13} Wang, R., Wagg, J., Carilli, C.~L., et al.\ 2013, \apj, 773, 44 
\bibitem[Willott et al.(2015)]{willott15} Willott, C.~J., Bergeron, J., \& Omont, A.\ 2015, \apj, 801, 123 
\bibitem[Willott et al.(2011)]{willott11} Willott, C.~J., Chet, S., Bergeron, J., \& Hutchings, J.~B.\ 2011, \aj, 142, 186 
\bibitem[Willott et al.(2005)]{willott05} Willott, C.~J., Delfosse, X., Forveille, T., Delorme, P., \& Gwyn, S.~D.~J.\ 2005, \apj, 633, 630 
\bibitem[Willott et al.(2007)]{willott07} Willott, C.~J., Delorme, P., Omont, A., et al.\ 2007, \aj, 134, 2435 
\bibitem[Willott et al.(2010a)]{willott10a} Willott, C.~J., Albert, L., Arzoumanian, D., et al.\ 2010a, \aj, 140, 546 
\bibitem[Willott et al.(2010b)]{willott10} Willott, C.~J., Delorme, P., Reyl{\'e}, C., et al.\ 2010b, \aj, 139, 906 
\bibitem[Willott et al.(2009)]{willott09} Willott, C.~J., Delorme, P., Reyl{\'e}, C., et al.\ 2009, \aj, 137, 3541 
\bibitem[Willott et al.(2013)]{willott13} Willott, C.~J., Omont, A., \& Bergeron, J.\ 2013, \apj, 770, 13 
\bibitem[Wu et al.(2015)]{wu15} Wu, X.-B., Wang, F., Fan, X., et al.\ 2015, \nat, 518, 512 
\bibitem[York et al.(2000)]{york00} York, D.~G., Adelman, J., Anderson, J.~E., Jr., et al.\ 2000, \aj, 120, 1579 
\bibitem[Zeimann et al.(2011)]{zeimann11} Zeimann, G.~R., White, R.~L., Becker, R.~H., et al.\ 2011, \apj, 736, 57 
\end{thebibliography}
\end{document}